\documentclass[acmsmall,screen,review=false, nonacm]{acmart}

\usepackage{algorithmic}
\usepackage{algorithm}
\usepackage{threeparttable}
\usepackage{multirow}
\usepackage{subfig,graphicx}

\usepackage{amsmath}
\usepackage{cases}

\AtBeginDocument{%
  \providecommand\BibTeX{{%
    \normalfont B\kern-0.5em{\scshape i\kern-0.25em b}\kern-0.8em\TeX}}}

\setcopyright{acmcopyright}
\copyrightyear{2018}
\acmYear{2018}
\acmDOI{10.1145/1122445.1122456}

\acmJournal{JACM}
\acmVolume{37}
\acmNumber{4}
\acmArticle{111}
\acmMonth{8}

\begin{document}

\title{Communication Efficient Federated Learning with Adaptive Quantization}

\author{Yuzhu Mao}
\affiliation{%
	\institution{Tsinghua Shenzhen International Graduate School, Tsinghua University}
	\city{Shenzhen}
	\country{China}
	\postcode{518055}
}
\affiliation{%
	\institution{Tsinghua-Berkeley Shenzhen Institute (TBSI), Tsinghua University}
	\city{Shenzhen}
	\country{China}
	\postcode{518055}
}
\email{myz20@mails.tsinghua.edu.cn}

\author{Zihao Zhao}
\affiliation{%
	\institution{Tsinghua Shenzhen International Graduate School, Tsinghua University}
	\city{Shenzhen}
	\country{China}
	\postcode{518055}
}
\affiliation{%
	\institution{Tsinghua-Berkeley Shenzhen Institute (TBSI), Tsinghua University}
	\city{Shenzhen}
	\country{China}
	\postcode{518055}
}
\email{kevinzhaozh1998@gmail.com}

\author{Guangfeng Yan}
\affiliation{%
	\institution{City University of Hong Kong, Hong Kong}
	\country{China}
}
\affiliation{%
	\institution{City University of Hong Kong Shenzhen Research Institute}
	\city{Shenzhen}
	\country{China}
}
\email{gfyan2-c@my.cityu.edu.hk}

\author{Yang Liu}
\affiliation{%
	\institution{Department of AI, WeBank}
	\city{Shenzhen}
	\country{China}
}
\email{yangliu@webank.com}

\author{Tian Lan}
\affiliation{%
	\department{Department of Electrical and Computer Engineering}
	\institution{George Washington University}
	\city{DC}
	\country{USA}
}
\email{tlan@gwu.edu}

\author{Linqi Song}
\affiliation{%
	\institution{City University of Hong Kong, Hong Kong}
	\country{China}
}
\affiliation{%
	\institution{City University of Hong Kong Shenzhen Research Institute}
	\city{Shenzhen}
	\country{China}
}
\email{linqi.song@cityu.edu.hk}

\author{Wenbo Ding$^*$}
\affiliation{%
	\institution{Tsinghua Shenzhen International Graduate School, Tsinghua University}
	\city{Shenzhen}
	\country{China}
	\postcode{518055}
}
\affiliation{%
	\institution{Tsinghua-Berkeley Shenzhen Institute (TBSI), Tsinghua University}
	\city{Shenzhen}
	\country{China}
	\postcode{518055}
}
\email{ding.wenbo@sz.tsinghua.edu.cn}


\begin{abstract}
 Federated learning (FL) has attracted tremendous attentions in recent years due to its privacy preserving measures and great potentials in some distributed but privacy-sensitive applications like finance and health. However, high communication overloads for transmitting high-dimensional networks and extra security masks remains a bottleneck of FL. This paper proposes a communication-efficient FL framework with Adaptive Quantized Gradient (AQG) which adaptively adjusts the quantization level based on local gradient’s update to fully utilize the heterogeneousness of local data distribution for reducing unnecessary transmissions. Besides, the client dropout issues are taken into account and the Augmented AQG is developed, which could limit the dropout noise with an appropriate amplification mechanism for transmitted gradients. Theoretical analysis and experiment results show that the proposed AQG leads to $25\%$-$50\%$ of additional transmission reduction as compared to existing popular methods including Quantized Gradient Descent (QGD) and Lazily Aggregated Quantized (LAQ) gradient-based method without deteriorating convergence properties. Particularly, experiments with heterogenous data distributions corroborate a more significant transmission reduction compared with independent identical data distributions. Meanwhile, the proposed AQG is robust to a client dropping rate up to 90$\%$ empirically, and the Augmented AQG manages to further improve the FL system’s communication efficiency with the presence of moderate-scale client dropouts commonly seen in practical FL scenarios.
\end{abstract}



\maketitle

\section{Introduction}
\label{sec:Introduction}
The deployment of Internet of things (IoT), ubiquitous sensing, edge computing and many other distributed systems have enabled the fast development of distributed learning techniques in recent years\cite{vehicularIoT, liu2020fedvision, Googlekeyboard}. The distributed learning could fully utilize the low-cost computing resources throughout the network and achieve comparable performance with the centralized learning. Nevertheless, the leakage of the data, gradient, and even model during the updating and transmitting process in distributed learning has raised the concerns of user privacy and security, which greatly limit its applications in some specific fields, such as finance, health, and etc. To this end, the federated learning (FL) which prevents privacy leakage by avoiding data exposition has been proposed by Google and other researchers, and attracted tremendous attentions from both academia and industry\cite{IEEEexample:1}.

Many approaches like differential privacy\cite{DifferentialPrivacy}, secret sharing techniques\cite{GoogleSecure} and homomorphic encryption\cite{liu2020secure} have been developed to mask the transmitted gradients and can almost well address the security issues in FL. However, high-dimensional neural networks and extra security masks\cite{ liu2019Concept, liu2019Secureboost, secure2020nature} may lead to high communication overhead, which becomes a main bottleneck of FL systems. In this context, the communication-efficient learning algorithms have been proposed mainly to reduce the transmission bits based on gradient quantization, which maps a real-valued vector to a constant number of bits. Representative gradient quantization algorithms for distributed systems include the Quantized Stochastic Gradient Descent (QSGD)\cite{QSGD}, 1-bit SGD\cite{1bit} and SignSGD\cite{signSGD}, etc. However, these methods communicate at all iterations (transmit all computed gradients) with a fixed number of quantization bits, which is not efficient enough for FL where non-IID (Independently Identically Distributed) data distribution is common. To address this problem, Sun \textit{et al.} proposed a gradient innovation-based Lazily Aggregated Quantized (LAQ) gradient method, which utilizes the differences between local loss functions and skips the transmission of slowly-varying quantized gradients\cite{LAQTPAMI}. Although LAQ reduces transmission overload by skipping unnecessary communication rounds, it still fixes the number of bits for all transmitted gradients, which remains to be improved.

In order to further reduce overall transmitted bits, this paper proposes a communication efficient FL framework with \textbf{A}daptive \textbf{Q}uantized \textbf{G}radient (AQG), where the quantization level is adjusted according to the local gradient’s updates adaptively. Specifically, gradients with larger amount of updates are quantized and transmitted with more bits, and vice versa. Besides, this paper takes client dropouts into account, which is another main challenge faced by FL system due to limited device reliability\cite{GoogleSecure}. In order to improve the performance of AQG with the presence of the noise introduced by client dropouts, the proposed FL framework with AQG is augmented by a variance-reduced method, where transmitted gradients are appropriately amplified to keep the unbiased estimators. 

Theoretical analysis and experiment results show that the proposed AQG outperforms existing methods in terms of overall transmitted bits without deteriorating convergence properties. Meanwhile, AQG is robust to a client dropping rate up to $90\%$ empirically, and the Augmented AQG with gradient amplification does act as a competitive solution to achieve an even more significant transmission reduction with moderate clients dropping scale commonly seen in practical FL scenarios.

The remainder of the paper is organized as follows. Section~\ref{sec:Motivations} provides the FL system overviews and discusses our motivations. The proposed Adaptive Quantized Gradient method is elaborated in section~\ref{sec:3a}. Theoretical analysis and convergence guarantee of AQG are provided in section~\ref{convergence}. We evaluate the performance of AQG with extensive experiments in section~\ref{sec:Evaluation} and conclude this paper in section~\ref{sec:conclusion}.

\textbf{Notation.} The notations involved in this paper are listed in Table~\ref{tab:notation}.

\renewcommand\arraystretch{1.5}
\begin{table}[h]
	\footnotesize
	\caption{Notations}
	\label{tab:notation}
	\centering
	\begin{tabular}{l | l}
		\hline
		\hline
		$\boldsymbol{g}_{m}^{k}$ & gradient computed by client $m$ at iteratoin $k$                               \\ \hline
		$\hat{\boldsymbol{g}}_{m}^{k}$ & gradient used for aggregation from client $m$ at iteration $k$            \\ \hline
		${b}_{max}$       & upper bound for the number of bits after quantization                                      \\ \hline
		$b_{m}^{k}$   & the quantization bit number chosen by client $m$ at iteration $k$ \\ \hline
		$\hat{b}_{m}^{k}$ & the quantization bit number chosen by client $m$ for $\hat{\boldsymbol{g}}_{m}^{k}$   \\ \hline
		$Q_{b}(\boldsymbol{g}_{m}^{k})$ & $\boldsymbol{g}_{m}^{k}$ quantized with $b$ bits                                                                 \\ \hline
		$\boldsymbol{\theta}^{k}$       & the aggregated global model broadcasted at iteration $k$                                                               \\ \hline
		$\varepsilon_{b}({\boldsymbol{g}}_{m}^{k})$    & quantization error $(Q_{b}(\boldsymbol{g}_{m}^{k})-\boldsymbol{g}_{m}^{k})$  \\ \hline
		$\mathbb{M}$ & clients set\\ \hline
		$\mathbb{M}_{b}^{k}$ & subset of clients uploading gradients with $b$ bits at iteration $k$\\ \hline
		$p$ & clients dropping rate  \\  \hline         
		$\lceil a \rceil$ & the ceil of $a$     \\                 \hline
		$\left \| \textbf{x} \right \|_{2}$ & $l_{2}$-norm of $\textbf{x}$ \\ \hline
		$\left \| \textbf{x} \right \|_{\infty}$ & $l_{\infty}$-norm of $\textbf{x}$ \\
		\hline
		\hline
	\end{tabular}
\end{table}

\section{System overview and Motivations}
\label{sec:Motivations}

\subsection{Federated Learning System}
\begin{figure}[htbp]
	\centering
	\includegraphics[width=0.7\textwidth]{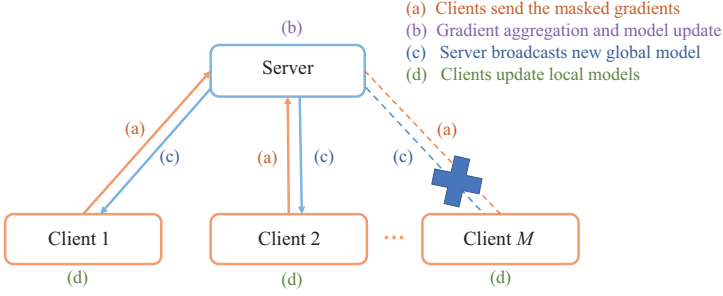}
	\caption{Typical architecture for a FL system.}
	\label{fig:FL}
\end{figure}

FL is designed to collaboratively train a global machine learning model with heterogeneous local data distribution across multiple privacy-sensitive clients. A typical architecture for a FL system with $M$ distributed clients and a server is shown in Fig.~\ref{fig:FL}. Similar to most distributed learning systems, FL system uses a server to receive locally-computed gradients and update global model by aggregation. However, in order to prevent privacy leakage from raw gradients, distributed clients have to mask or encrypt the local gradients before transmission. Therefore, the communication burden in FL systems tends to be heavier compared with other distributed learning systems\cite{GoogleSecure}. Besides, distributed clients in FL systems, such as mobile devices in wireless networks, usually have limited computation and communication resources, which may lead to the dropout of the participants in each iteration, like the client $M$ shown in Fig.~\ref{fig:FL}. Thus, the robustness to client dropout is another practical requirement for FL systems\cite{GoogleSecure}.

\subsection{Motivations}
FL is bottlenecked by the high communication overheads and limited device reliability. The lack of efficient transmission and robustness to client dropouts may lead to slow, expensive and unstable learning. In this paper, the FL framework with the proposed AQG method provides opportunities for communication-efficient FL with large-scale of client dropouts.  

Firstly, AQG focuses on reducing unnecessary transmission by fully utilizing the heterogeneous property of FL. Due to the heterogeneousness of local data distribution, local optimization objectives descend at different rates. Therefore, adaptively adjusting the quantization level according to gradient’s update amount provides a more efficient way to communicate with the server by quantizing slowly-varying gradients with less amount of bits.

Secondly, AQG aims to address the noise induced by client dropouts. When a client dropout occurs, all coordinates of transmitted gradient are lost, which can be regarded as an extreme example of gradient sparsification\cite{Sparsification1, Sparsification2, Sparsification3, Sparsification4}. In order to limit the variance increase of a sparsified gradient, Wangni \textit{et al.} proposed to keep the unbiasedness of the sparsified gradient by appropriately amplifying the remaining coordinates\cite{Amplification}. Inspired by this idea, AQG tries to stay robust to client dropouts or even further improve the communication efficiency of FL with client dropouts by further adjusting the transmitted gradients and suppressing the noise.

\section{AQG: Adaptive Quantized Gradient}
\label{sec:3a}
\begin{figure}[htbp]
	\centering
	\includegraphics[width=0.75\textwidth]{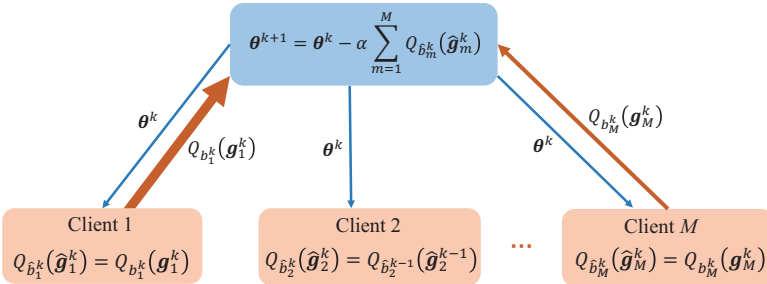}
	\caption{FL with AQG.}
	\label{fig:AQG}
\end{figure}

To reduce the transmission overheads, a multilevel adaptive quantization scheme is proposed in this section. As illustrated in Fig.~\ref{fig:AQG}, the FL system with AQG can be implemented as follows. At iteration $k$, the server broadcasts global model $\boldsymbol{\theta} ^{k}$ to all clients. Each client computes gradient $\boldsymbol{g}_{m}^{k}$ by taking all its local data $\textbf{X}_{m}$ as a full batch:
\begin{equation}
\label{eq1}
\begin{aligned}
\boldsymbol{g}_{m}^{k}&=\nabla f_{m}(\textbf{X}_{m};\boldsymbol{\theta} ^{k}) 
\end{aligned}
\end{equation}

After the gradient computation, each client needs to make \textbf{two decisions}: (1) is it necessary to send its quantized gradient? (2) how many bits $b_{m}^{k}$ should be used to quantize and send its newly-computed gradient? In particular, the first decision is the key idea in LAQ\cite{LAQTPAMI}. In this paper, it is considered as a special case of the second decision, where $b_{m}^{k}$ is chosen as zero if the client decides to send nothing.

If client $m$ chooses a non-zero $b_{m}^{k}$ and updates its newly-quantized gradient, then $Q_{b_{m}^{k}}(\boldsymbol{g}_{m}^{k})$ is one of the quantized gradients that actually participate in gradient aggregation on the server side at iteration $k$. Otherwise, the server reuses the old quantized-gradient $Q_{\hat{b}_{m}^{k-1}}(\hat{\boldsymbol{g}}_{m}^{k-1})$ from the last iteration to represent client $m$ in the aggregation. In summary, an iteration step of proposed AQG is as follows:

\begin{equation}
\textbf{Gradients Update} \quad \quad \quad \quad \quad \quad  Q_{\hat{b}_{m}^{k}}(\hat{\boldsymbol{g}}_{m}^{k})=\left\{
\begin{aligned}
&Q_{b_{m}^{k}}(\boldsymbol{g}_{m}^{k}),\quad m \in \mathbb{M}\setminus \mathbb{M}_{0}^{k}\\
&Q_{\hat{b}_{m}^{k-1}}(\hat{\boldsymbol{g}}_{m}^{k-1}),\quad m \in \mathbb{M}_{0}^{k}
\end{aligned}
\quad \quad \quad \quad \quad \quad \quad \quad \right.
\end{equation}

\begin{equation}
\textbf{Gradients Aggregation} \quad \quad \quad \quad   \boldsymbol{\theta} ^{k+1} = \boldsymbol{\theta} ^{k} - \alpha\sum_{m \in \mathbb{M}}Q_{\hat{b}_{m}^{k}}(\hat{\boldsymbol{g}}_{m}^{k})
\quad \quad \quad \quad \quad \quad \quad \quad \quad \quad \quad \quad
\end{equation}
where $\mathbb{M}_{0}^{k}$ denotes the subset of clients that sets $b_{m}^{k}=0$ and uploads nothing at iteration $k$. For client $m$, $Q_{\hat{b}_{m}^{k}}(\hat{\boldsymbol{g}}_{m}^{k})$ represents the quantized gradient actually used for aggregation at iteration $k$, which may be outdated if $m \in \mathbb{M}_{0}^{k}$. 

The target problems of AQG is that: 
\begin{itemize}
	\item [1)] 
	For clients belonging to $\mathbb{M}\setminus \mathbb{M}_{0}^{k}$, the precision levels (quantization levels) of their new updates $Q_{b_{m}^{k}}(\boldsymbol{g}_{m}^{k})$ are not fixed, but adaptively adjusted depending on $\boldsymbol{g}_{m}^{k}$'s innovations——the difference between the newly-quantized gradient and the last quantized gradient sent to the server. It motivates a need for not only a quantization scheme as previous work, but also a \textbf{precision selection criterion} to decide the quantization level of each newly-computed gradient;
	\item [2)]  
	For FL scenario where client dropouts is relatively frequent, methods to limit the noise introduced by gradients lossing are also in great need.
\end{itemize}

The following part of this section presents the precision selection criterion developed in this paper and the quantization scheme applied in the proposed AQG. At last, an optional augmentation of AQG is proposed to address potential client dropouts.

\subsection{Precision Selection Criterion}

As mentioned before, the LAQ algorithm proposed by Sun \textit{et al.} skips the uploads of 
quantized gradients with small innovations——the difference between $Q_{b}(\boldsymbol{g}_{m}^{k})$ and the last upload $Q_{b}(\hat{\boldsymbol{g}}_{m}^{k-1})$, where $b$ is the fixed number of bits after quantization\cite{LAQTPAMI}. In order to decide whether client $m$ needs to upload its newly-quantized gradient $Q_{b}(\boldsymbol{g}_{m}^{k})$ at iteration $k$, LAQ develops a communication selection criterion as follows:
\begin{align}
\left \| Q_{b}(\hat{\boldsymbol{g}}_{m}^{k-1})-Q_{b}(\boldsymbol{g}_{m}^{k}) \right \|_{2}^{2} \geq \frac{1}{\alpha ^{2}M^{2}}\sum_{d=1}^{D}\xi _{d}\left \| \boldsymbol{\theta} ^{k+1-d}-\boldsymbol{\theta} ^{k-d} \right \|_{2}^{2} +3(\left \| \varepsilon _{b}(\hat{\boldsymbol{g}}_{m}^{k-1})\right \|_{2}^{2}+\left \| \varepsilon _{b}({\boldsymbol{g}}_{m}^{k}) \right \|_{2}^{2}) \label{eq4}
\end{align}
where $\varepsilon _{b}(\hat{\boldsymbol{g}}_{m}^{k-1})$ and $\varepsilon _{b}({\boldsymbol{g}}_{m}^{k})$ denote quantization errors, and $\left \{ \xi_{d} \right \}_{d=1}^{D}$ are predetermined constant weights used to balance the impact of global model updates from previous $D$ steps. In LAQ, client $m$ sends its newly-quantized local gradient $Q_{b}(\boldsymbol{g}_{m}^{k})$ at iteration $k$ only when the difference between $Q_{b}(\boldsymbol{g}_{m}^{k})$ and the last upload $Q_{b}(\hat{\boldsymbol{g}}_{m}^{k-1})$ is larger than a threshold, which takes the quantization error and global model's innovation into account\cite{LAQTPAMI}.

This paper extends the single precision level LAQ with communication selection criterion \eqref{eq4} to multilevel adaptive quantization for transmitted gradients. The key idea of AQG is that under a pre-set upper bound $b_{max}$ for the number of bits after 
quantization, gradients with smaller innovations can be quantized with less number of bits, since the negative impact of their precision losses on convergence is limited.

In order to decide how many bits $b_{m}^{k}$ should be used to quantize and send client $m$'s newly-computed gradient $\boldsymbol{g}_{m}^{k}$, we develop the following precision selection criterion: 
\begin{align}
&\left \| Q_{\hat{b}_{m}^{k-1}}(\hat{\boldsymbol{g}}_{m}^{k-1})-Q_{b_{max}}(\boldsymbol{g}_{m}^{k}) \right \|_{2}^{2} \geq \nonumber  \\
&\frac{1}{\alpha ^{2}M^{2}}\sum_{d=1}^{D}\xi _{d}\left \| \boldsymbol{\theta} ^{k+1-d}-\boldsymbol{\theta} ^{k-d} \right \|_{2}^{2} +3(\left \| \varepsilon _{b_{max}-b+1}(\hat{\boldsymbol{g}}_{m}^{k-1})\right \|_{2}^{2}+\left \| \varepsilon _{b_{max}-b+1}({\boldsymbol{g}}_{m}^{k}) \right \|_{2}^{2}) \label{eq6}
\end{align}

\begin{figure}[h]
	\centering
	\includegraphics[width=0.7\textwidth]{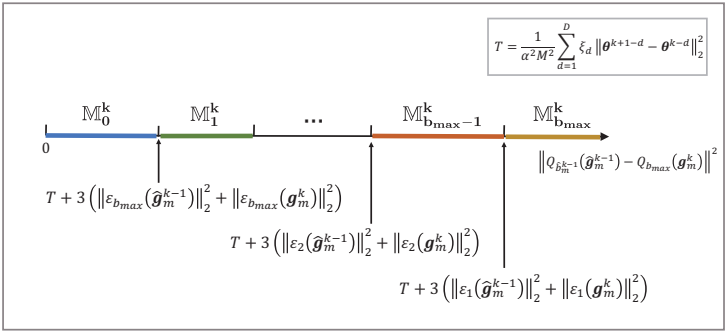}
	\caption{The principle of the precision selection criterion.}
	\label{fig:class}
\end{figure}

As illustrated in Fig.~\ref{fig:class}, the proposed precision selection criterion \eqref{eq6} works in the following ways:
\begin{itemize}
	\item [1)] 
	For any $\bar{b} \in [1,..., b_{max}-1]$, satisfying \eqref{eq6} with $b=\bar{b}+1$ will necessarily satisfy \eqref{eq6} with $b=\bar{b}$, but not vice versa. The reason is that for a given ${\boldsymbol{g}}_{m}^{k}$, there is always $\varepsilon _{b_{max}-(\bar{b}+1)+1}({\boldsymbol{g}}_{m}^{k}) = \varepsilon _{b_{max}-\bar{b}}({\boldsymbol{g}}_{m}^{k}) \geq  \varepsilon _{b_{max}-\bar{b}+1}({\boldsymbol{g}}_{m}^{k})$ due to more error brought by more aggressive quantization.  
	\item [2)]
	Precision selection criterion \eqref{eq6} with $b=1$ acts as communication selection criterion in AQG. Specifically, if \eqref{eq6} with $b=1$ does not hold for client $m$, then its gradient update at iteration $k$ is skipped.
\end{itemize}

Therefore, client subsets devided by the proposed precision criterion form the client set $\mathbb{M}$ without overlaps:
\begin{equation}
\mathbb{M}_{0}^{k} \cup \mathbb{M}_{1}^{k}\cup \mathbb{M}_{2}^{k}\cup ...\cup \mathbb{M}_{b_{max}}^{k} = \mathbb{M} 
\label{eq7}
\end{equation}
where $\mathbb{M}_{b}^{k}$ denotes the subset of clients which send gradients quantized
by $b$ bits at iteration $k$. In particular, $\mathbb{M}_{0}^{k}$ denotes the subset of clients which skip the update.

The FL with AQG is summarized in \textbf{Algorithm 1}. At iteration $k$, each client checks where its innovation locates in Fig.~\ref{fig:class}, and then re-quantizes its gradient with corresponding number of bits for update. Theoretical analysis of multilevel AQG with \eqref{eq6} is provided in section~\ref{convergence}.

For computation simplicity, a two-level variant of AQG is also proposed in this paper. At each iteration: \\
\textbf{Two-level AQG.} there are only two precision-levels to be selected for each client. In other words, $b$ in criterion \eqref{eq6} only has two options: $\lceil \frac{b_{max}}{2} \rceil$ and $b_{max}$.

\begin{center} 
	\begin{minipage}{.6\linewidth}
		\label{alg:AQG}
		\renewcommand{\algorithmicrequire}{\textbf{Input:}}
		\renewcommand{\algorithmicensure}{\textbf{Initialize:}}
		\begin{algorithm}[H]
			\caption{AQG}
			\begin{algorithmic}[1]
				\REQUIRE stepsize $\alpha > 0$, $b_{max}$, $D$, and $\{\xi _{d} \}_{d=1}^{D}$.
				\ENSURE $\boldsymbol{\theta} ^{1}$.
				\FOR {$k = 1,2,...,K$}
				\STATE Server broadcasts $\boldsymbol{\theta} ^{k}$ to all workers.
				\FOR {each client $m\in\mathbb{M}$ \textbf{in parallel}}
				\STATE Worker $m$ computes ${\boldsymbol{g}}_{m}^{k}$ and $Q_{b_{max}}({\boldsymbol{g}}_{m}^{k})$.
				\IF {\eqref{eq6} with $b=1$ holds for worker $m$}
				\FOR {$b = b_{max},b_{max}-1,...,1$}
				\IF {\eqref{eq6} with $b$ holds for worker $m$}
				\STATE Worker $m$ computes and sends $Q_{b}({\boldsymbol{g}}_{m}^{k})$.
				\STATE Set $b_{m}^{k}=b$.
				\STATE Set $\hat{\boldsymbol{g}}_{m}^{k} = \boldsymbol{g}_{m}^{k}$ and $\hat{b}_{m}^{k} = b$ on both sides.
				\STATE \textbf{Break.}
				\ENDIF
				\ENDFOR
				\ELSE
				\STATE Worker $m$ sends nothing.
				\STATE Set $b_{m}^{k}=0$,
				\STATE Set $\hat{\boldsymbol{g}}_{m}^{k} = \hat{\boldsymbol{g}}_{m}^{k-1}$ and $\hat{b}_{m}^{k} = \hat{b}_{m}^{k-1}$ on both sides.
				\ENDIF
				\ENDFOR
				\STATE Server updates $\boldsymbol{\theta} ^{k+1}$ by $\boldsymbol{\theta} ^{k}-\alpha \sum_{m=1}^{M}Q_{\hat{b}_{m}^{k}}(\hat{\boldsymbol{g}}_{m}^{k})$.
				\ENDFOR
			\end{algorithmic}
		\end{algorithm}
	\end{minipage}
\end{center}

\subsection{Quantization Scheme}
For better comparison, we adapt the quantization scheme used in LAQ algorithm\cite{LAQTPAMI}. The scheme quantizes the difference between the new gradient $\boldsymbol{g}_{m}^{k}$ and the last quantized upload $Q_{\hat{b}_{m}^{k-1}}(\hat{\boldsymbol{g}}_{m}^{k-1})$:
\begin{equation}
\Delta = \boldsymbol{g}_{m}^{k}-Q_{\hat{b}_{m}^{k-1}}(\hat{\boldsymbol{g}}_{m}^{k-1})
\end{equation}

\begin{figure}[h]
	\centering
	\includegraphics[width=0.6\textwidth]{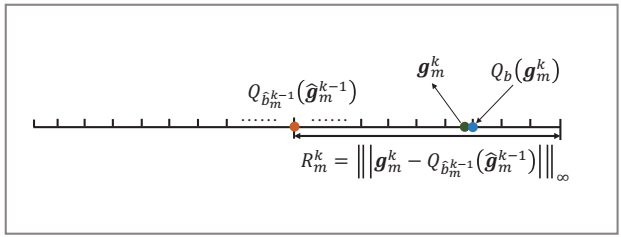}
	\caption{Quantization scheme in AQG.}
	\label{fig:quantization}
\end{figure}

With $b$ bits used for quantization, the value range of $\Delta$'s elements can be represented by a uniformly discretized grid with $2^{b}-1$ quantized values, as shown in Fig.~\ref{fig:quantization}. By projecting every real number in this range to the closest quantized value, $\boldsymbol{g}_{m}^{k}$ can be represented by $Q_{b}(\boldsymbol{g}_{m}^{k})$ with $b$ bits for each element instead of 32/64 bits by default.

\subsection{Augmented AQG for Client Dropouts}
This paper also considers random client dropout in FL, and uses $z_{m}^{k}$ to control the participation of client $m$ at iteration $k$. With a client dropping rate $p$:
\begin{equation*}
z_{m}^{k}\sim Bernoulli(p)
\label{eq8}
\end{equation*}

If $z_{m}^{k}=1$, client $m$ drops out and fails to perform gradient computation at iteration $k$. It is obvious that with a dropping rate $p$, the percentage of active clients is approximately $1-p$ at each iteration.

With such setting, the expectation of client $m$'s upload is as follows:
\begin{equation}
E[Q_{b_{m}^{k}}(\boldsymbol{g}_{m}^{k})] = (1-p)\cdot  Q_{b_{m}^{k}}(\boldsymbol{g}_{m}^{k}) + p\cdot \textbf{0}
\label{eq9}
\end{equation}
where $\textbf{0}$ is a zero vector of the same shape as $Q_{b_{m}^{k}}(\boldsymbol{g}_{m}^{k})$.

In order to get the unbaised expectation, the upload is adjusted to $Q_{b_{m}^{k}}(\boldsymbol{g}_{m}^{k})/(1-p)$, and then:
\begin{align}
E[Q_{b_{m}^{k}}(\boldsymbol{g}_{m}^{k})] = (1-p)\cdot  ( Q_{b_{m}^{k}}(\boldsymbol{g}_{m}^{k})/(1-p)) + p\cdot  \textbf{0}  = Q_{b_{m}^{k}}(\boldsymbol{g}_{m}^{k})
\label{Eq:14}
\end{align}

The Augmented AQG is summarized in \textbf{Algorithm 2}. The intuitive explanation for gradient amplification is that the loss function $f_{m}$ is smooth, which means the new update $ Q_{b_{m}^{k}}(\boldsymbol{g}_{m}^{k})$ tends to be approximate to recent previous updates that may have been lost due to client dropouts.  

\begin{center}
\begin{minipage}{.6\linewidth}
	\label{alg:AAQG}
	\renewcommand{\algorithmicrequire}{\textbf{Input:}}
	\renewcommand{\algorithmicensure}{\textbf{Initialize:}}
	\begin{algorithm}[H]
		\caption{Augmented AQG}
		\begin{algorithmic}[1]
			\REQUIRE stepsize $\alpha > 0$, $b_{max}$, $D$, and $\{\xi _{d} \}_{d=1}^{D}$.
			\ENSURE $\boldsymbol{\theta} ^{1}$.
			\FOR {$k = 1,2,...,K$}
			\STATE Server broadcasts $\boldsymbol{\theta} ^{k}$ to all workers.
			\FOR {each client $m\in\mathbb{M}$ \textbf{in parallel}}
			\IF{$z_{m}^{k} = 1$}
			
			\STATE Worker $m$ computes ${\boldsymbol{g}}_{m}^{k}$ and $Q_{b_{max}}({\boldsymbol{g}}_{m}^{k})$.
			\IF {\eqref{eq6} with $b=1$ holds for worker $m$}
			\FOR {$b = b_{max},b_{max}-1,...,1$}
			\IF {\eqref{eq6} with $b$ holds for worker $m$}
			\STATE Worker $m$ computes and sends $Q_{b}({\boldsymbol{g}}_{m}^{k})$.
			\STATE Set $b_{m}^{k}=b$.
			\STATE Set $\hat{\boldsymbol{g}}_{m}^{k} = \boldsymbol{g}_{m}^{k}$ and $\hat{b}_{m}^{k} = b$ on both sides.
			\STATE \textbf{Break.}
			\ENDIF
			\ENDFOR
			
			\ENDIF
			\ELSE
			
			\STATE Worker $m$ sends nothing.
			\STATE Set $b_{m}^{k}=0$,
			\STATE Set $\hat{\boldsymbol{g}}_{m}^{k} = \hat{\boldsymbol{g}}_{m}^{k-1}$ and $\hat{b}_{m}^{k} = \hat{b}_{m}^{k-1}$ on both sides.
			
			\ENDIF
			\ENDFOR
			\STATE Server updates $\boldsymbol{\theta} ^{k+1}$ by $\boldsymbol{\theta} ^{k}-\alpha \sum_{m=1}^{M}Q_{\hat{b}_{m}^{k}}(\hat{\boldsymbol{g}}_{m}^{k})$.
			\ENDFOR
		\end{algorithmic}
	\end{algorithm}
\end{minipage}
\end{center}

Compared to the existing LAQ method, the proposed AQG method adjusts the number of quantization bits based on local gradient innovation adaptively. The rationale of AQG is that the proposed precision selection criterion utilizes the inherent heterogeneousness of local optimization objectives to reduce unnecessary transmission cost. Theoretical analysis in the next section will prove that AQG maintains the desired convergence properties of LAQ. Experiments show that AQG advances and fits FL better with following contributions:
\begin{itemize}
	\item [1)] 
	AQG outperforms existing popular methods in terms of overall transmission bits, and achieves a more significant transmission reduction with heterogeneous data distribution compared to IID data distribution;
	\item [2)]
	AQG is robust to a clients dropping rate up to 90$\%$, and the Augmented AQG manages to further reduce transmission overload with the presence of moderate-scale of client dropouts.
\end{itemize}

\section{Convergence analysis}
\label{convergence}
In this section, the proposed AQG is analyzed theoretically and a convergence guarantee is provided. The theoretical analysis of AQG is based on following assumption:  \\
\textbf{Assumption 1.} \textit{Loss function $f(\boldsymbol{\theta} ) = \sum_{m\in \mathbb{M}}f_{m}(\boldsymbol{\theta} )$ is L-smooth.}

The Lyapunov function of AQG is defined in the same way as LAQ: 
\begin{equation}
\mathbb{V}(\boldsymbol{\theta} ^{k})=f(\boldsymbol{\theta} ^{k})-f(\boldsymbol{\theta} ^{*})+\sum_{d=1}^{D}\sum_{j=d}^{D}\frac{\xi  _{j}}{\alpha }\left \| \boldsymbol{\theta} ^{k+1-d}- \boldsymbol{\theta} ^{k-d} \right \|_{2}^{2} 
\label{eqV}
\end{equation}  
where $\boldsymbol{\theta} ^{*}$ is the optimal solution of $\min_{\boldsymbol{\theta}} f(\boldsymbol{\theta})$.

With the quantization errors in precision selection criterion \eqref{eq6} being ignored, the parameter differences term in Lyapunov function helps guarantee that the error induced by skipping gradients decreases with the objective residual in the training process.

\subsection{Convergence Guarantee}
\label{convergence_a}
To ensure convergence, the following inequality should always hold:
\begin{equation}
\mathbb{V}(\boldsymbol{\theta} ^{k+1})-\mathbb{V}(\boldsymbol{\theta} ^{k})\leq 0
\label{eq9}
\end{equation}
\textbf{Lemma 1.} \textit{Under Assumption 1, \eqref{eq9} holds if the following three inequalities are satisfied simultaneously:}
\begin{subequations}
	\begin{align}
	-\frac{\alpha }{2}+\frac{1}{2}\alpha \rho _{1}+(L+2\beta _{1})(1+\rho _{2})\alpha ^{2} \leq 0 \label{Za}\\
	[\frac{\alpha }{2}+(\frac{L}{2}+\beta _{1})(1+{\rho _{2}}^{-1})\alpha ^{2}]\frac{\xi _{D}}{\alpha^{2}}-\beta _{D} \leq 0 \label{Zb} \\
	[\frac{\alpha }{2}+(\frac{L}{2}+\beta _{1})(1+{\rho _{2}}^{-1})\alpha ^{2}]\frac{\xi _{d}}{\alpha^{2}}+\beta _{d+1}-\beta _{d} \leq 0 \label{Zc}
	\end{align}
	\label{eq:lemma12}
\end{subequations}
\textit{where $\rho _{1}$ and $\rho _{2}$ are constants. $\beta _{d}=\frac{1}{\alpha }\sum_{j=d}^{D}\xi _{j}, \forall d \in \{1,...,D\}$. See the appendix for proof details.}  \\

It indicates that if the stepsize $\alpha$ and constants $\left \{ \xi_{d} \right \}_{d=1}^{D}$ satisfy the three inequalities above, the convergence of the Lyapunov function \eqref{eqV} is guaranteed theoretically.

\subsection{Linear Convergence With Strongly-Convex Loss}
The theoretical analysis under strongly-convex loss function is based on the following assumption: \\
\textbf{Assumption 2.} \textit{Loss function $f(\boldsymbol{\theta} ) = \sum_{m\in \mathbb{M}}f_{m}(\boldsymbol{\theta} )$ is µ-strongly convex. }

Under Assumption 2, there is:
\begin{equation}
\left \| \boldsymbol{\theta} -\boldsymbol{\theta} ^{*} \right \|_{2}^{2}\leq \frac{2}{\mu }[f(\boldsymbol{\theta} )-f(\boldsymbol{\theta} ^{*})]
\label{eq11}
\end{equation} 
\textbf{Lemma 2.} \textit{Under Assumption 1 and 2, the following inequality holds:}
\begin{align}
&\mathbb{V}(\boldsymbol{\theta} ^{k+1}) \leq (1-c)\mathbb{V}(\boldsymbol{\theta} ^{k})  \nonumber \\
&+ B \left \| \sum_{m=1}^{M}\varepsilon_{b_{max}}(\hat{\boldsymbol{g}}_{m}^{k}) \right \|_{2}^{2} +B\sum_{m\in \mathbb{M}_{0}^{k}}(\left \| \varepsilon _{b_{max}}(\hat{\boldsymbol{g}}_{m}^{k-1})\right \|_{2}^{2}+\left \| \varepsilon _{b_{max}}({\boldsymbol{g}}_{m}^{k}) \right \|_{2}^{2}) \nonumber \\
&+B( \sum_{b=1}^{b_{max}}\sum_{m\in \mathbb{M}_{b}^{k}} \left \| \varepsilon _{\hat{b}_{m}^{k}}(\hat{\boldsymbol{g}}_{m}^{k}) \right \|_{2}^{2} +  \sum_{b=1}^{b_{max}}\sum_{m\in \mathbb{M}_{b}^{k}} \left \| \varepsilon _{b_{max}}(\hat{\boldsymbol{g}}_{m}^{k}) \right \|_{2}^{2}) 
\label{eq12}
\end{align}
\textit{where $c$ and $B$ are constants depending on $\mu$, $\rho _{1}$, $\rho _{2} $ and parameters involved in selection criterion \eqref{eq6}. See the appendix for proof details.}  \\
\textbf{Theorem 1.} \textit{Under Assumption 1, Assumption 2 and Lemma 2, Lyapunov function and the quantization errors all converge at a linear rate:}
\begin{subequations}
	\label{L}
	\begin{align}
	\left \| \varepsilon _{b}(\boldsymbol{g}_{m} ^{k}) \right \|_{\infty}^{2} &\leq P{\tau_{b}}^{2} \sigma ^{k}\mathbb{V}(\boldsymbol{\theta} ^{1}) \label{La} \\
	\mathbb{V}(\boldsymbol{\theta} ^{k+1})&\leq \sigma ^{k}\mathbb{V}(\boldsymbol{\theta} ^{1}) \label{Lb}
	\end{align}
\end{subequations}
\textit{where $\sigma \in (0, 1)$ and $\tau_{b}$ is the quantization granularity with $2^{b}$ quantization levels. $P$ is a constant based on parameters in Lemma 1. See the appendix for proof details.}

\renewcommand\arraystretch{1.5}
\begin{table*}[h]
	\scriptsize
	\centering
	\begin{threeparttable}
		\caption{Performance comparison of gradient-based algorithms.}
		\label{tb:1}
		\begin{tabular}{c|c|c|c|c|c|c}
			\toprule[1.5pt]
			\multicolumn{3}{c|}{\textbf{Experiment setting}}                                         & \textbf{Iteration \#} & \textbf{Communication \#} & \textbf{Bit \#} & \textbf{Transmission Reduction} \\ \hline
			\hline
			\multirow{8}{*}{Logistic Regression} & \multirow{4}{*}{IID}     & \textbf{Two Level AQG}  & 500          & 3933             & $\boldsymbol{7952}$ & $\boldsymbol{41\%}$  \\ \cline{3-7} 
			&                          & \textbf{Multilevel AQG} & 500          & 4372            & $\boldsymbol{8372}$  & $\boldsymbol{38\%}$ \\ \cline{3-7} 
			&                          & 4-bit LAQ       & 500          & 3354             & $1.34\times 10^{4}$ & 0 \\ \cline{3-7} 
			&                          & 4-bit QGD       & 500          & 9000             & $3.6\times 10^{4}$ & $-$ \tnote{*} \\ \cline{2-7} 
			& \multirow{4}{*}{non-IID} & \textbf{Two Level AQG}  & 500         & 4870            & $\boldsymbol{1.54\times 10^{4}}$ & $\boldsymbol{51\%}$  \\ \cline{3-7} 
			&                          & \textbf{Multilevel AQG} & 500         & 8273            & $\boldsymbol{1.78\times 10^{4}}$ & $\boldsymbol{43\%}$ \\ \cline{3-7} 
			&                          & 4-bit LAQ       & 500         & 7842            & $3.14\times 10^{4}$  & 0 \\ \cline{3-7} 
			&                          & 32-bit GD \tnote{1}       & 500         & 9000           & $2.88\times 10^{5}$ & $-$ \\ \hline
			\multirow{8}{*}{Neural Network}      & \multirow{4}{*}{IID}     & \textbf{Two Level AQG}  & 2713         & 854              & $\boldsymbol{1708}$ & $\boldsymbol{34\%}$  \\ \cline{3-7} 
			&                          & \textbf{Multilevel AQG} & 2881         & 974              & $\boldsymbol{1928}$  & $\boldsymbol{25\%}$ \\ \cline{3-7} 
			&                          & 4-bit LAQ       & 2784         & 643              & 2572  & 0  \\ \cline{3-7} 
			&                          & 4-bit QGD       & 2890         & 28900        & $1.16\times 10^{5}$ & $-$ \\ \cline{2-7} 
			& \multirow{4}{*}{non-IID} & \textbf{Two Level AQG}  & 1319         & 1030              & $\boldsymbol{2060}$   & $\boldsymbol{44\%}$  \\ \cline{3-7} 
			&                          & \textbf{Multilevel AQG} & 1702         & 977              & $\boldsymbol{1845}$ & $\boldsymbol{49\%}$  \\ \cline{3-7} 
			&                          & 4-bit LAQ       & 2219         & 921              & 3684  & 0  \\ \cline{3-7} 
			&                          & 4-bit QGD       & 1251         & 12510            & 50040 & $-$ \\ 
			\bottomrule[1.5pt]
		\end{tabular}
		\begin{tablenotes}    
			\footnotesize               
			\item[1] Since 4-bit QGD fails to converge with logistic regression and non-IID data distribution, the 32-bit vanilla GD is implemented for comparison.   
			\item[*] 4-bit QGD definitely costs more bits compared against the baseline 4-bit LAQ. 
		\end{tablenotes}            
	\end{threeparttable}  
\end{table*}

\begin{figure}[htbp]
	\begin{minipage}[b]{.325\linewidth}
		\centering
		\subfloat[][Loss v.s. iteration]{
			\label{1a}
			\includegraphics[width=1\linewidth]{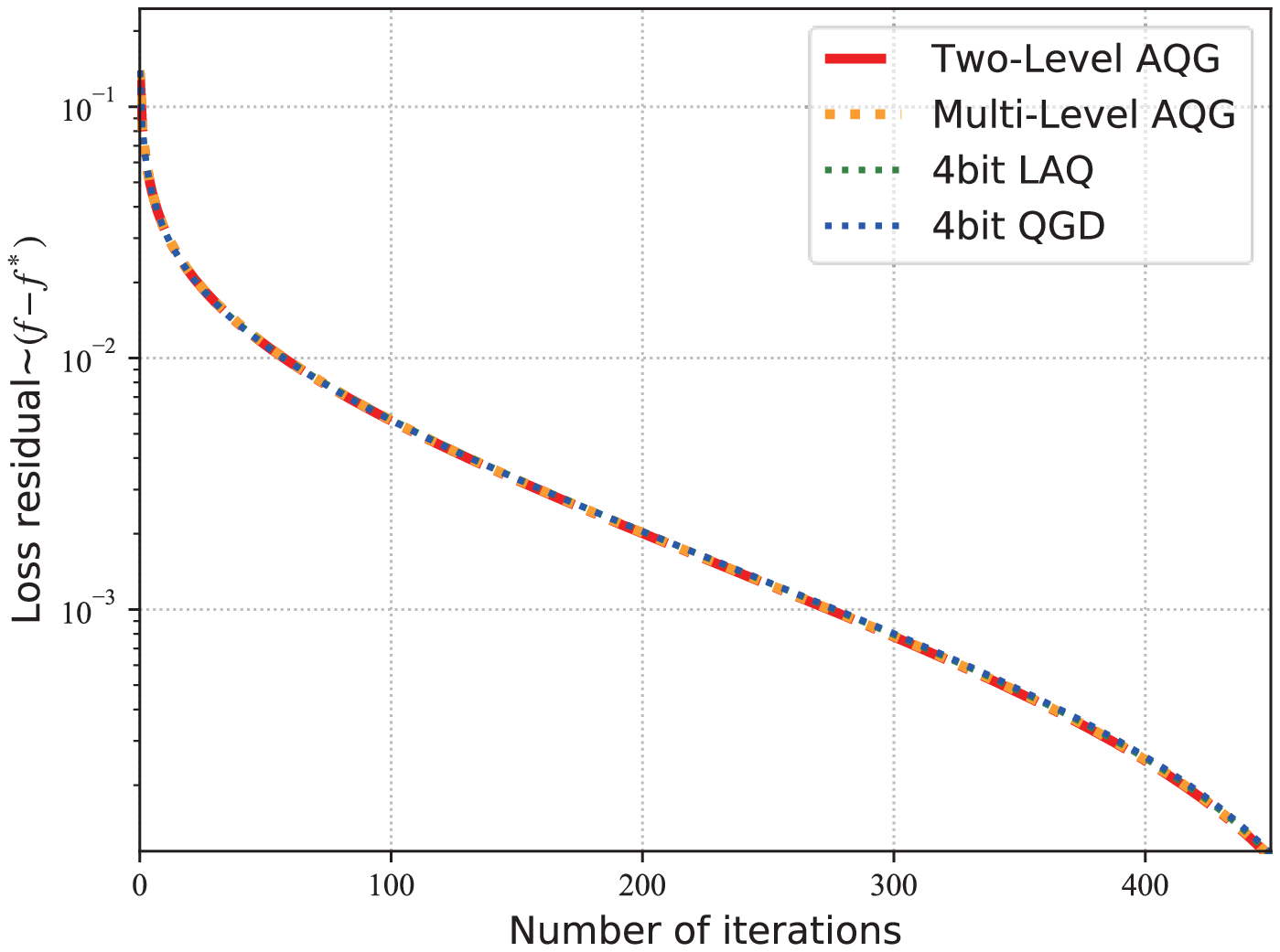}}
	\end{minipage} 
	\begin{minipage}[b]{.325\linewidth}
		\subfloat[][Loss v.s. communication round]{
			\label{1b}
			\includegraphics[width=1\linewidth]{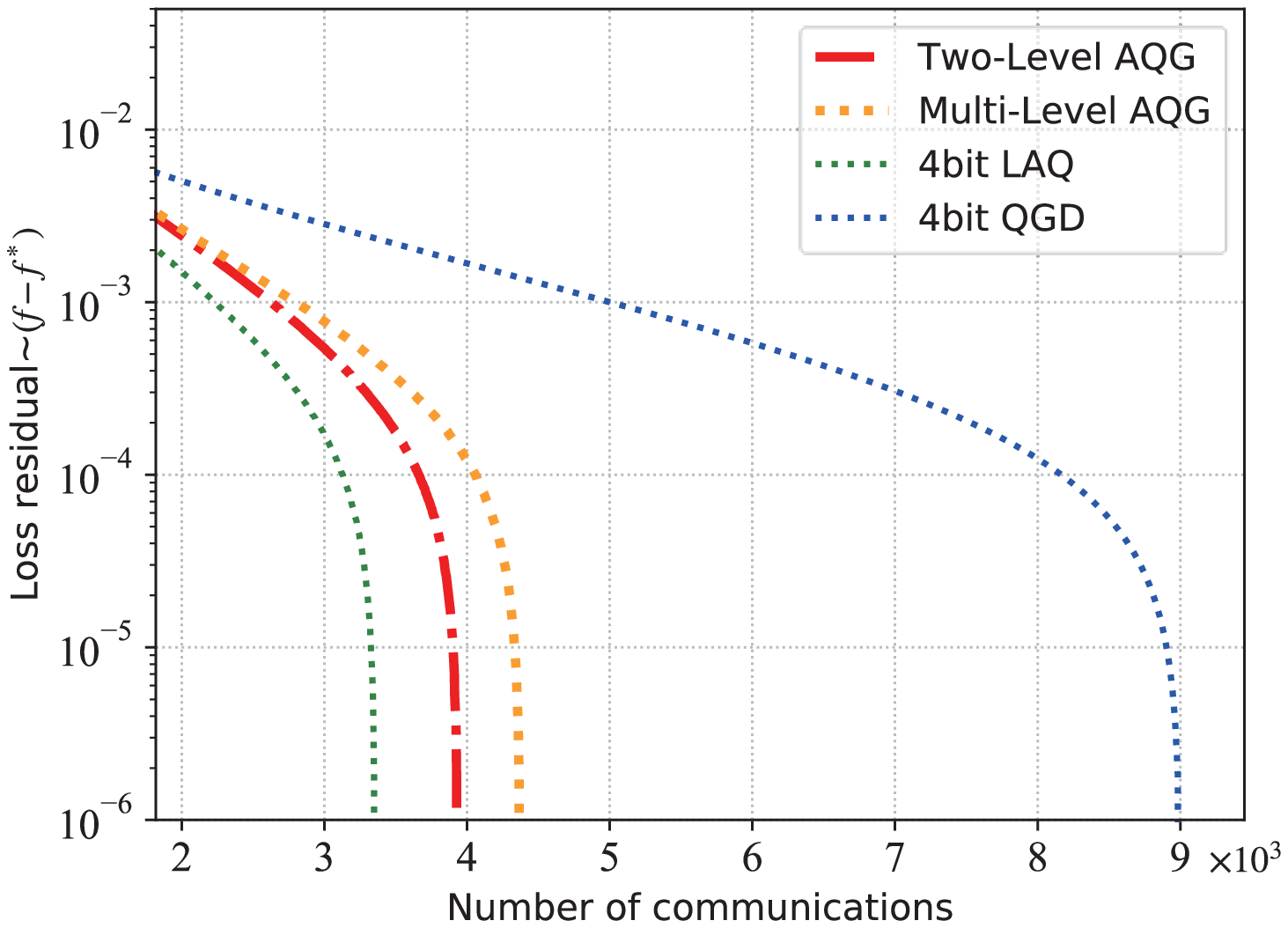}}
	\end{minipage} 
	\begin{minipage}[b]{.325\linewidth}
		\subfloat[][Loss v.s. bit]{
			\label{1c}
			\includegraphics[width=1\linewidth]{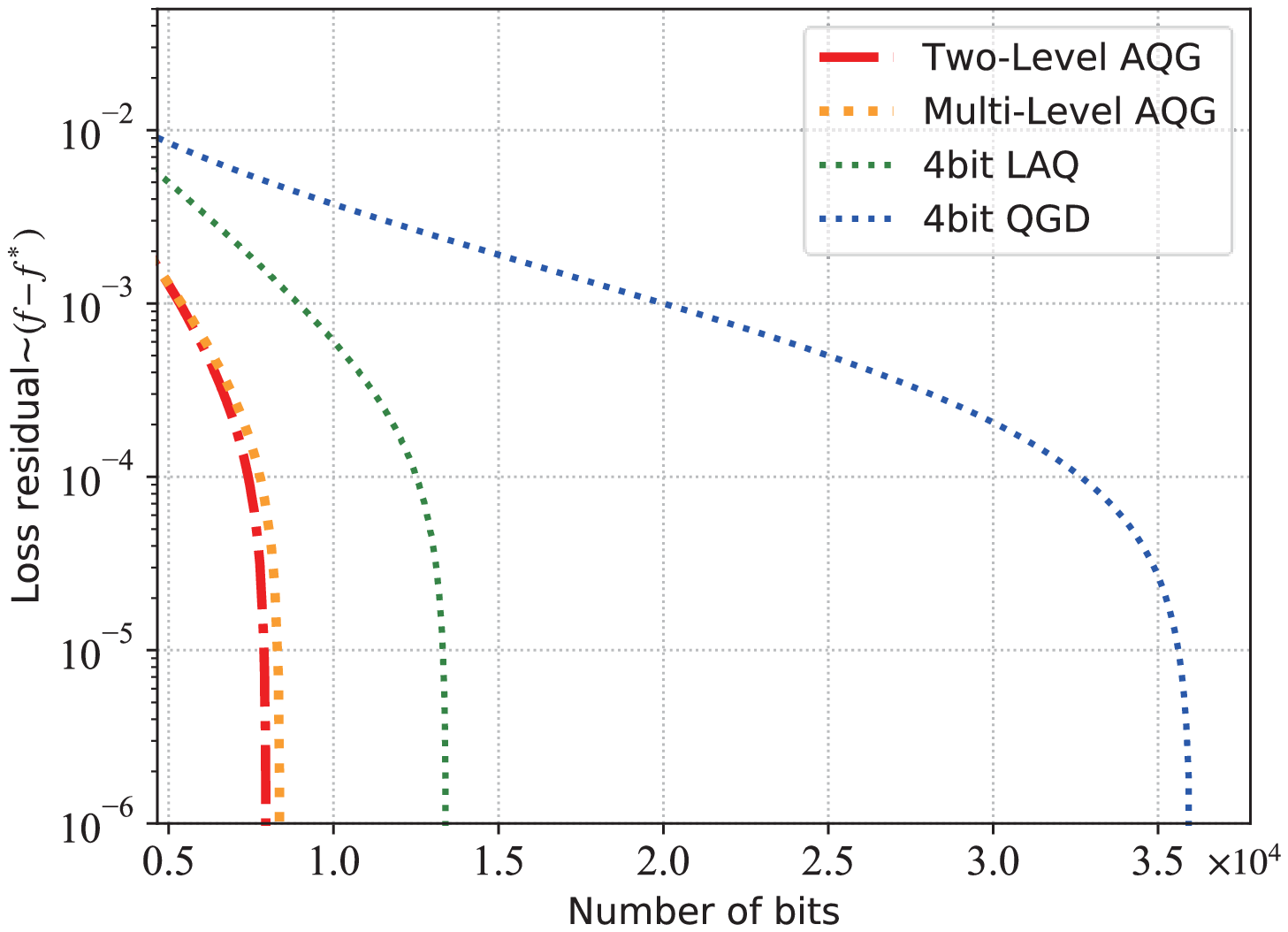}}
	\end{minipage} 
	\caption{Convergence of loss function with logistic regression and IID data distribution}
	\label{fig:logistic IID}
\end{figure}

\begin{figure}[htbp]
	\begin{minipage}[b]{.325\linewidth}
		\centering
		\subfloat[][Loss v.s. iteration]{
			\label{2a}
			\includegraphics[width=1\linewidth]{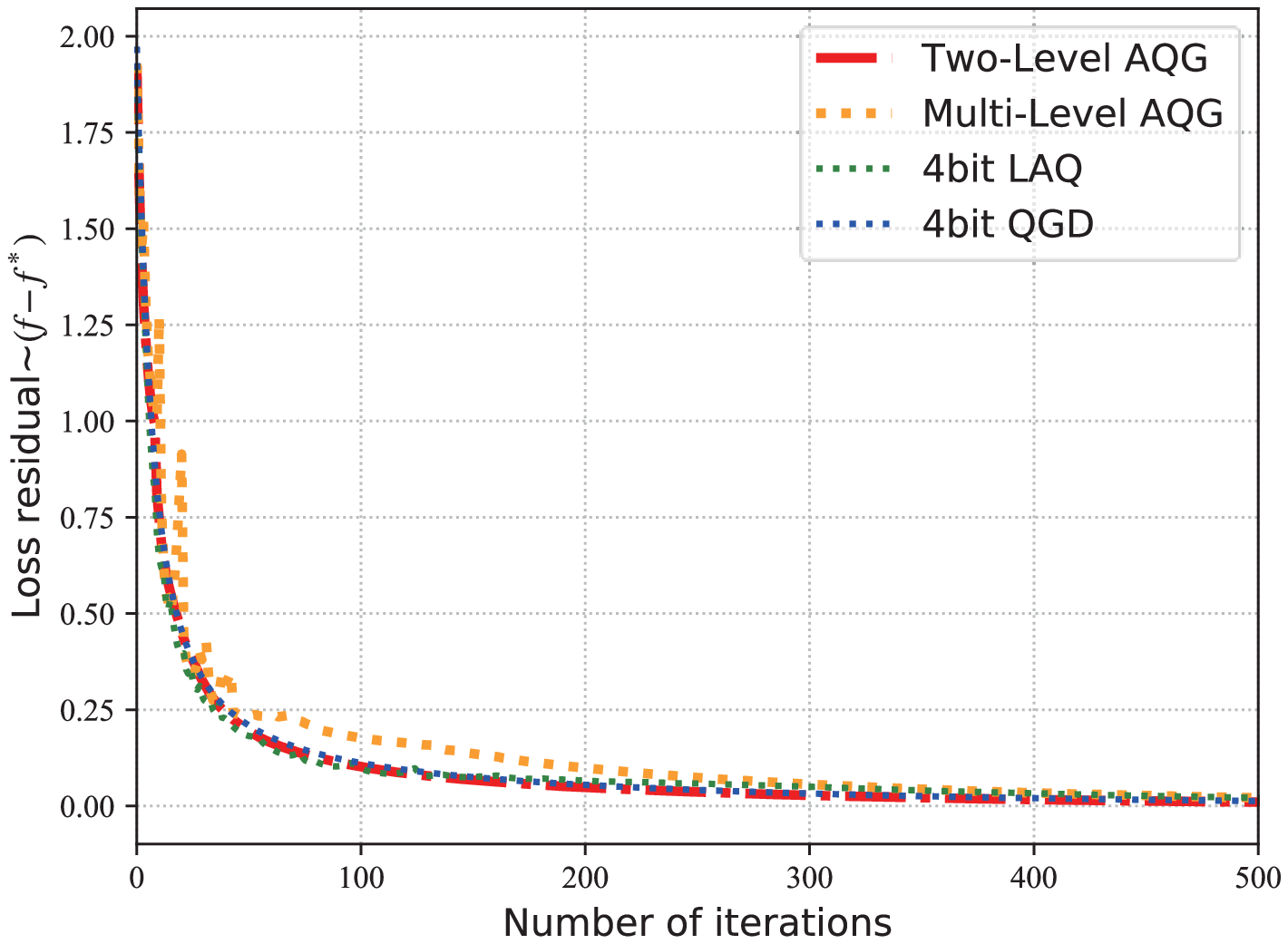}}
	\end{minipage} 
	\begin{minipage}[b]{.325\linewidth}
		\subfloat[][Loss v.s. communication round]{
			\label{2b}
			\includegraphics[width=1\linewidth]{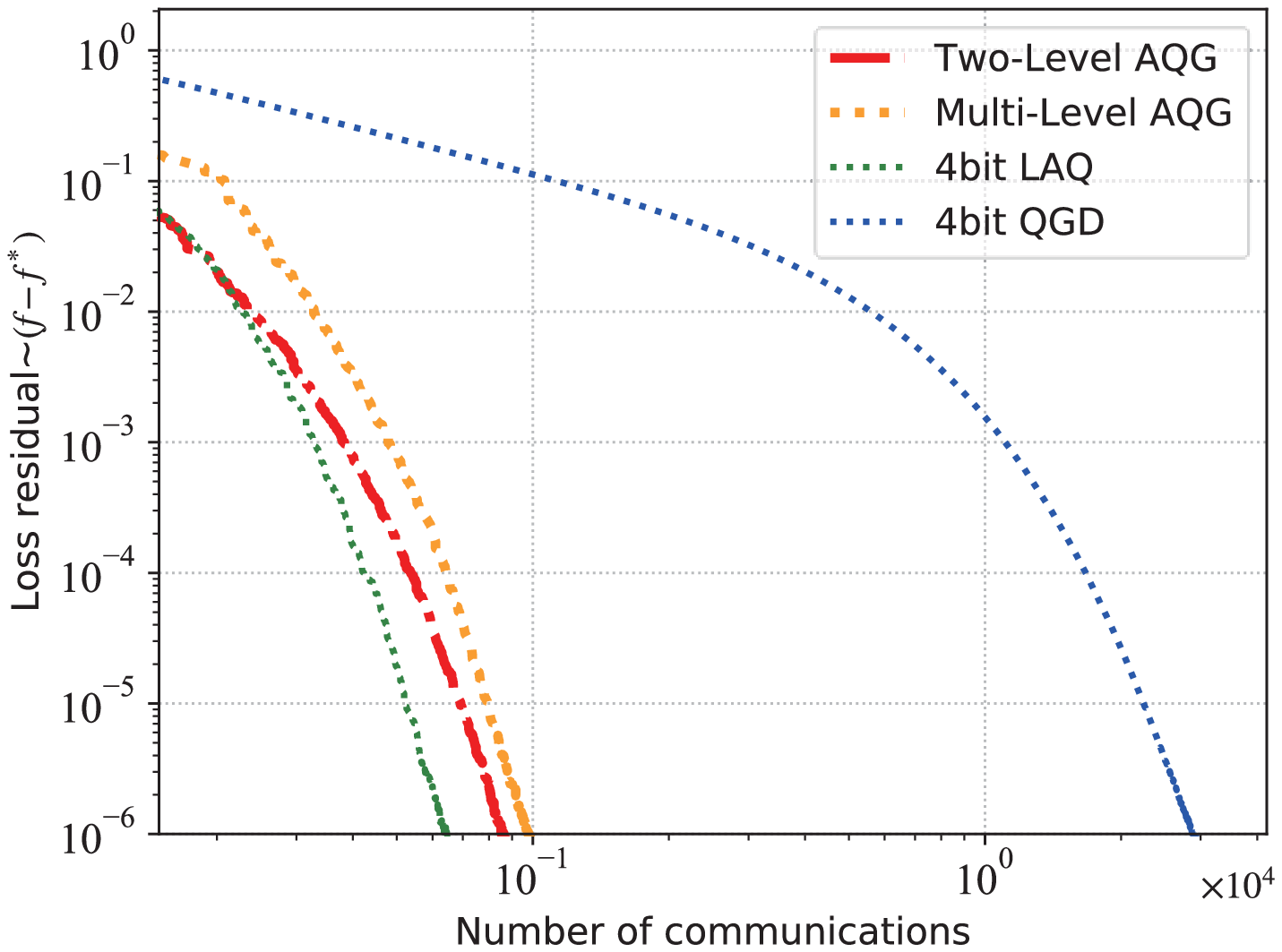}}
	\end{minipage} 
	\begin{minipage}[b]{.325\linewidth}
		\subfloat[][Loss v.s. bit]{
			\label{2c}
			\includegraphics[width=1\linewidth]{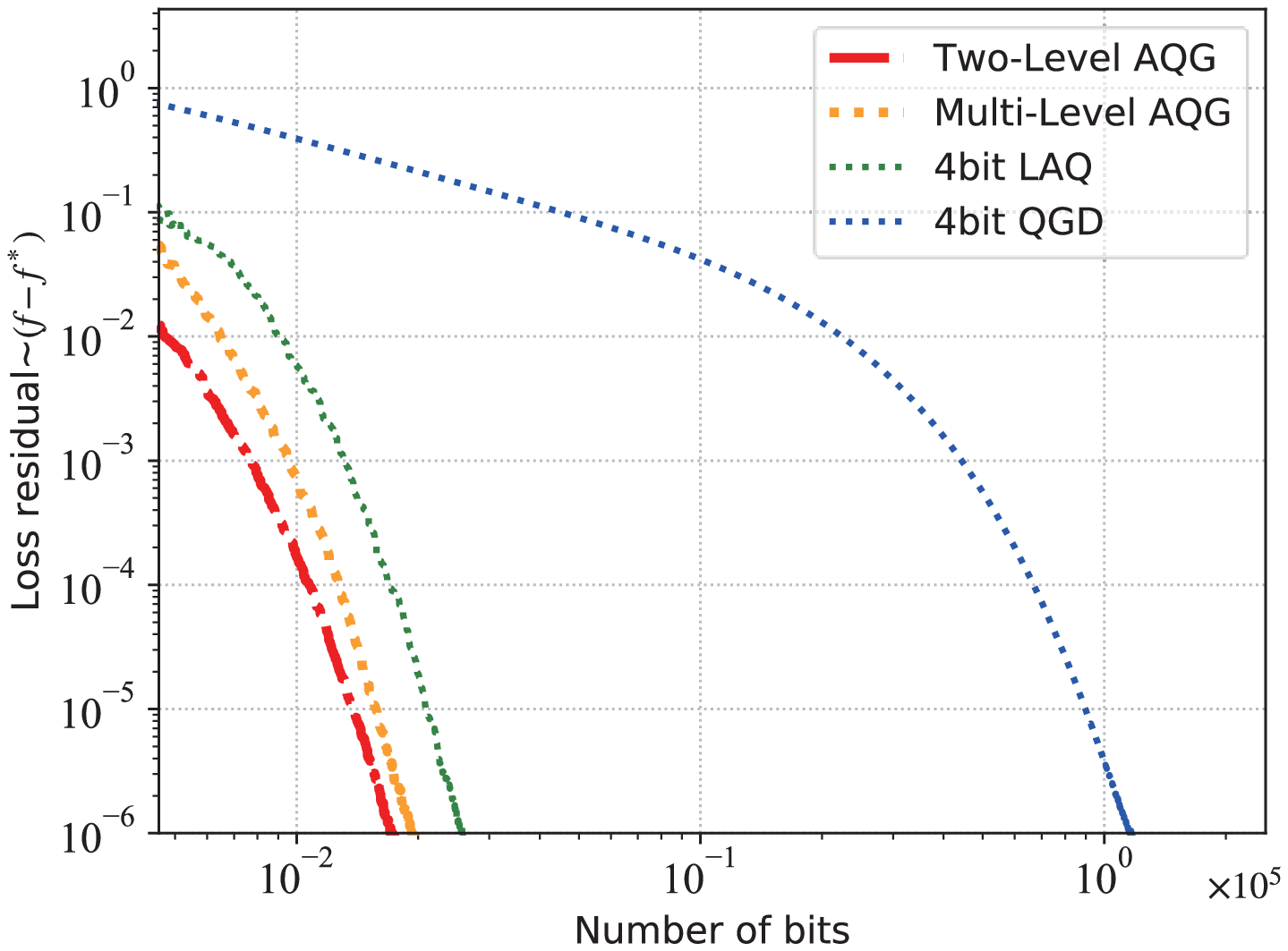}}
	\end{minipage} 
	\caption{Convergence of loss function with neural network and IID data distribution}
	\label{fig:network IID}
\end{figure}

\section{Experiment Results}
\label{sec:Evaluation}
In this section, the performance of FL with the proposed AQG is evaluated with regularized logistic regression and neural network, respectively representing strongly convex and non-convex loss function. Experiment results demonstrate that AQG outperforms state-of-the-art quantization algorithms including QGD and LAQ in terms of reducing transmission bits and resisting client dropouts.
\subsection{Experimental Settings}
For experiment simplicity, logistic regression is implemented with binary classification, and a fully connected network is built for non-convex optimization. The input and output dimension of the fully connected network is 784 and 10, respectively. For both Multi-level AQG and Tow-level AQG, the quantization bit number's upper bound $b_{max}$ is 4, the constant parameter $D$ is 10, and the weights $\left \{ \xi_{d} \right \}_{d=1}^{D}=1/D$. Stepsize $\alpha$ is 0.008 for logistic regression and 0.02 for neural network.  

In terms of datasets, both non-IID data distribution and IID data distribution are considered as follows: 

\textbf{non-IID Data Distribution:} To simulate non-IID data distribution, a heterogeneous simulation dataset including 18 distributed data slices is used for logistic regression, and MNIST Dataset is used for multi-classification with the fully connected network by assigning each client with only one class of samples. The detailed description of the adopted dataset is provided in appendix. Obviously, the total client number $M$ is set as 18 for logistic regression and 10 for fully connected network with these two datasets.

\textbf{IID Data Distribution:} For better comparison, the same binary classification dataset used to simulate non-IID data distribution is applied to simulate IID data distribution by uniformly distributing the samples across 18 clients. For the task with fully connected network, the MNIST dataset is distributed uniformly across 10 clients. Other parameters keep the same as in non-IID data distribution.

The experiment results are shown in Table.~\ref{tb:1}. For logistic regression, all algorithms run 500 iterations. For neural network, all algorithms run 4000 iterations, and we calculate the number of iteration, communication round and transmission bit when the loss residual decreases to less than $1\times 10^{-6}$. For both tasks, the amount of bits counted for each algorithm in Table.~\ref{tb:1} is the number of bits used to transmit \textbf{one} dimension of the uploaded gradient. Thus, the higher the dimension of gradient is, the more significant transmission reduction AQG brings.

\subsection{Performance of AQG with IID Data Distribution}
With IID data distribution, training samples are distributed uniformly among clients.  Fig.~\ref{1a} shows that Multi-level AQG and the two-level variant of AQG both reach linear convergence rate as LAQ and QGD in strongly convex condition. Meanwhile, AQG significantly saves transmission bits compared against 4-bit LAQ and 4-bit QGD, as shown in Fig.~\ref{1c}. It can be observed from Fig.~\ref{1b} that the reduction of transmission bits is at the cost of a slight increase in communication rounds compared with LAQ, but it is worthy due to the significant reduction in overall transmission load.

Fig.~\ref{fig:network IID} shows the results with non-convex loss function. Similar to the results with logistic regression, Multi-level AQG and two-level AQG both require fewer amount of bits to reach convergence without sacrificing the convergence properties of 4-bit LAQ and 4-bit QGD, as depicted in Fig.~\ref{2a} and Fig.~\ref{2c}. Meanwhile, compared with 4-bit QGD, AQG significantly reduces communication rounds to the same order of magnitude as 4-bit LAQ, as shown in Fig.~\ref{2b}.

\begin{figure}[htbp]
	\begin{minipage}[b]{.325\linewidth}
		\centering
		\subfloat[][Loss v.s. iteration]{
			\label{3a}
			\includegraphics[width=1\linewidth]{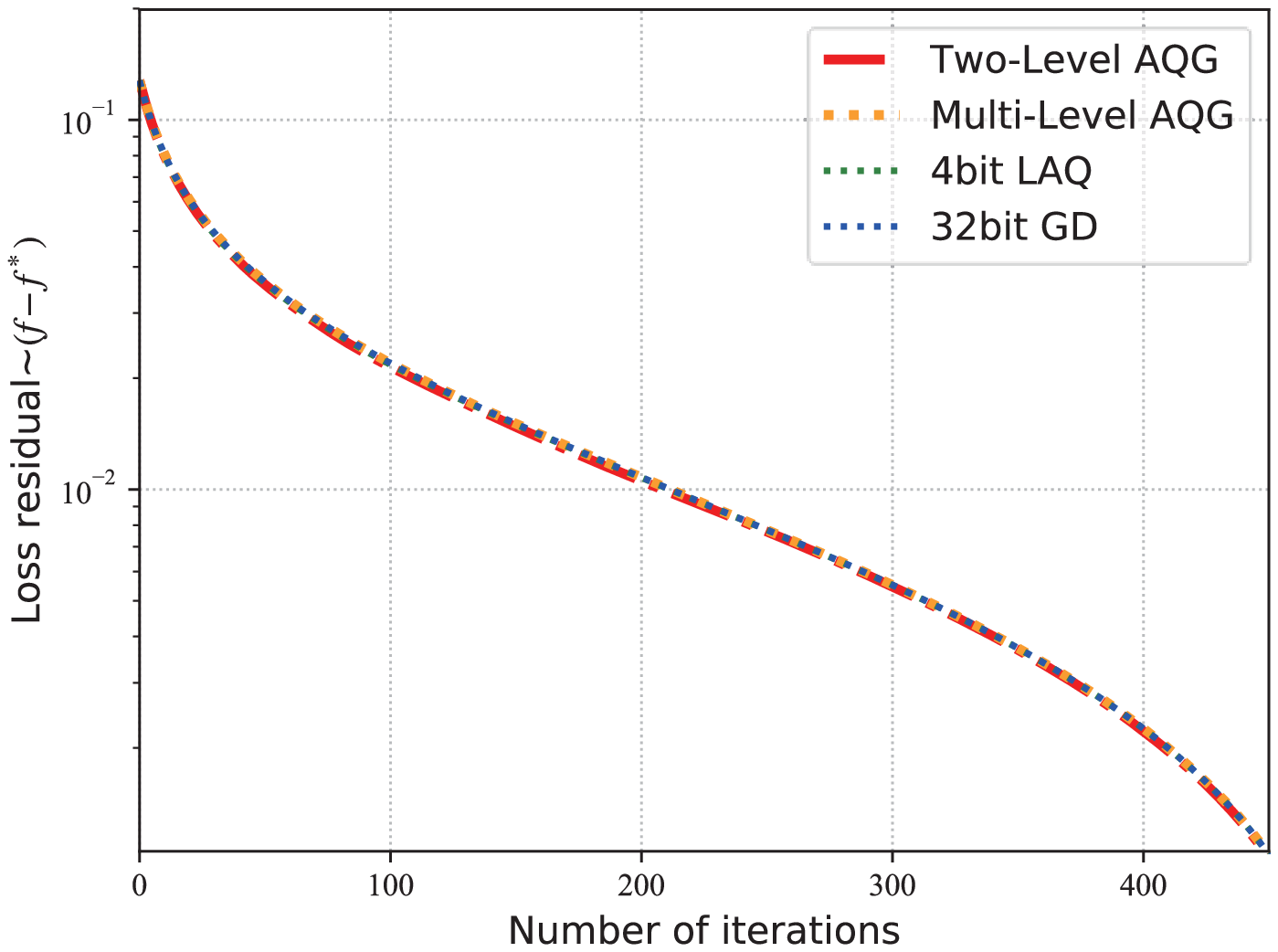}}
	\end{minipage} 
	\begin{minipage}[b]{.325\linewidth}
		\subfloat[][Loss v.s. communication round]{
			\label{3b}
			\includegraphics[width=1\linewidth]{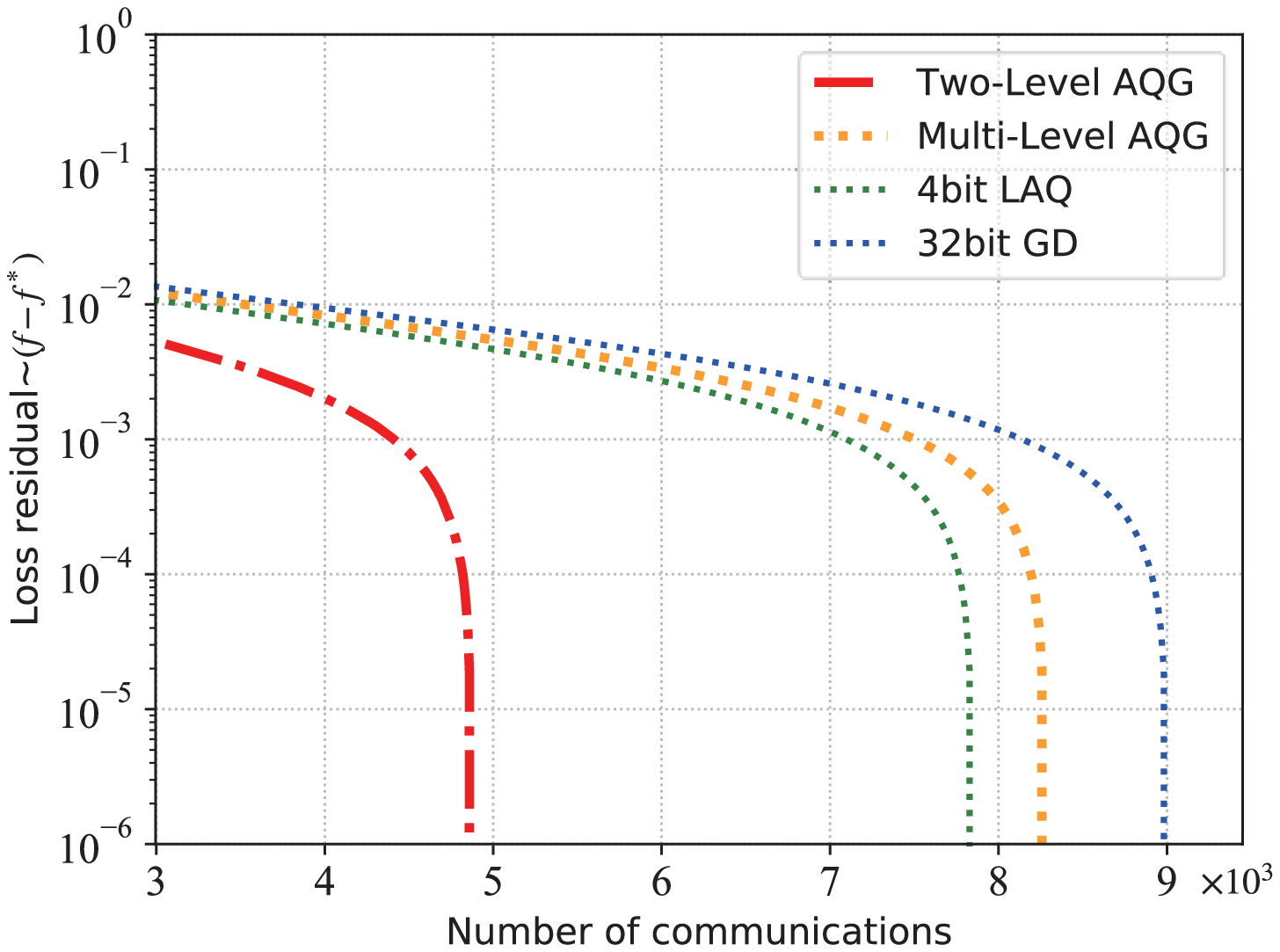}}
	\end{minipage} 
	\begin{minipage}[b]{.325\linewidth}
		\subfloat[][Loss v.s. bit]{
			\label{3c}
			\includegraphics[width=1\linewidth]{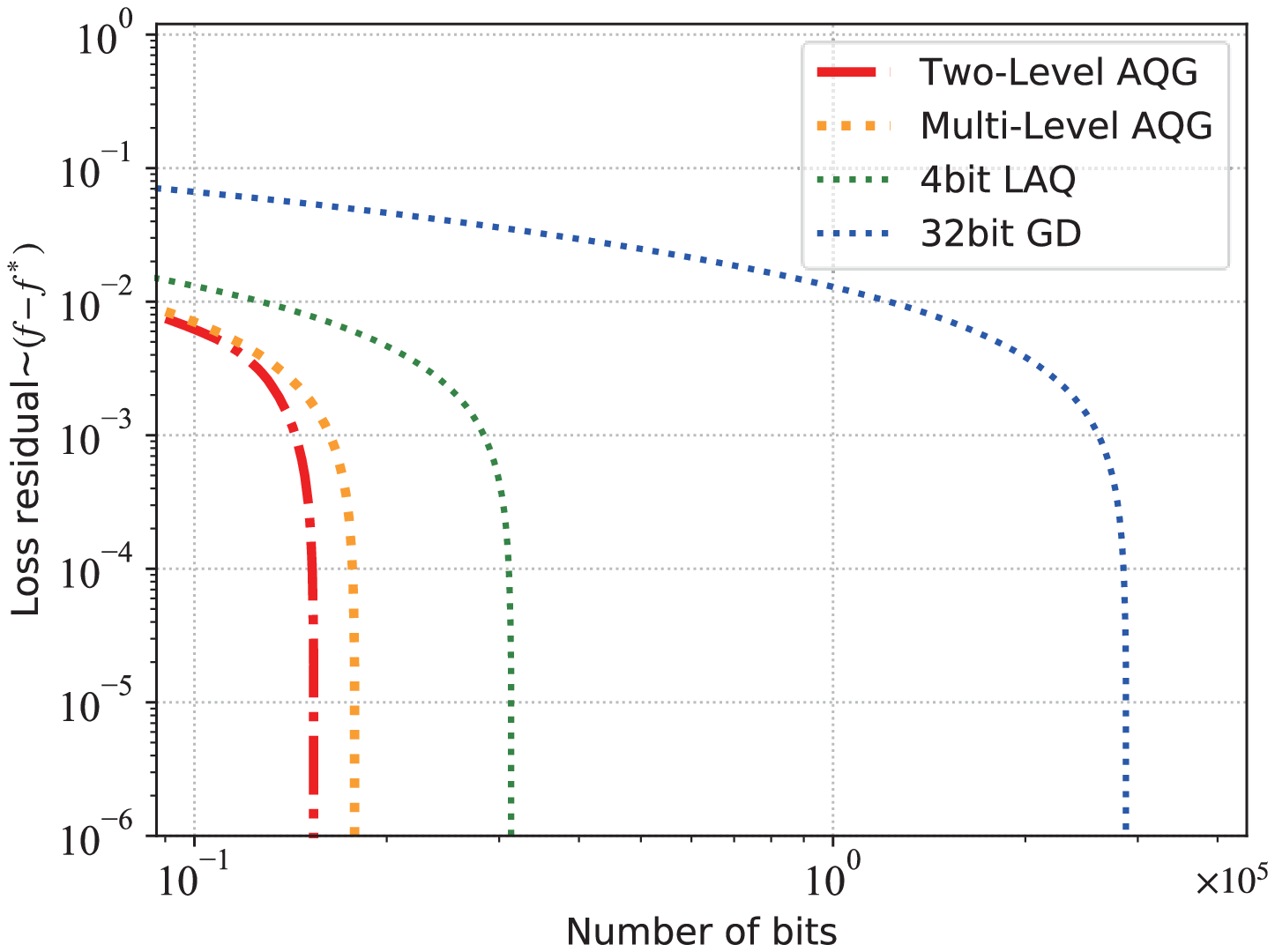}}
	\end{minipage} 
	\caption{Convergence of loss function with logistic regression and non-IID data distribution}
	\label{fig:logistic NonIID}
\end{figure}

\begin{figure}[htbp]
	\begin{minipage}[b]{.325\linewidth}
		\centering
		\subfloat[][Loss v.s. iteration]{
			\label{4a}
			\includegraphics[width=1\linewidth]{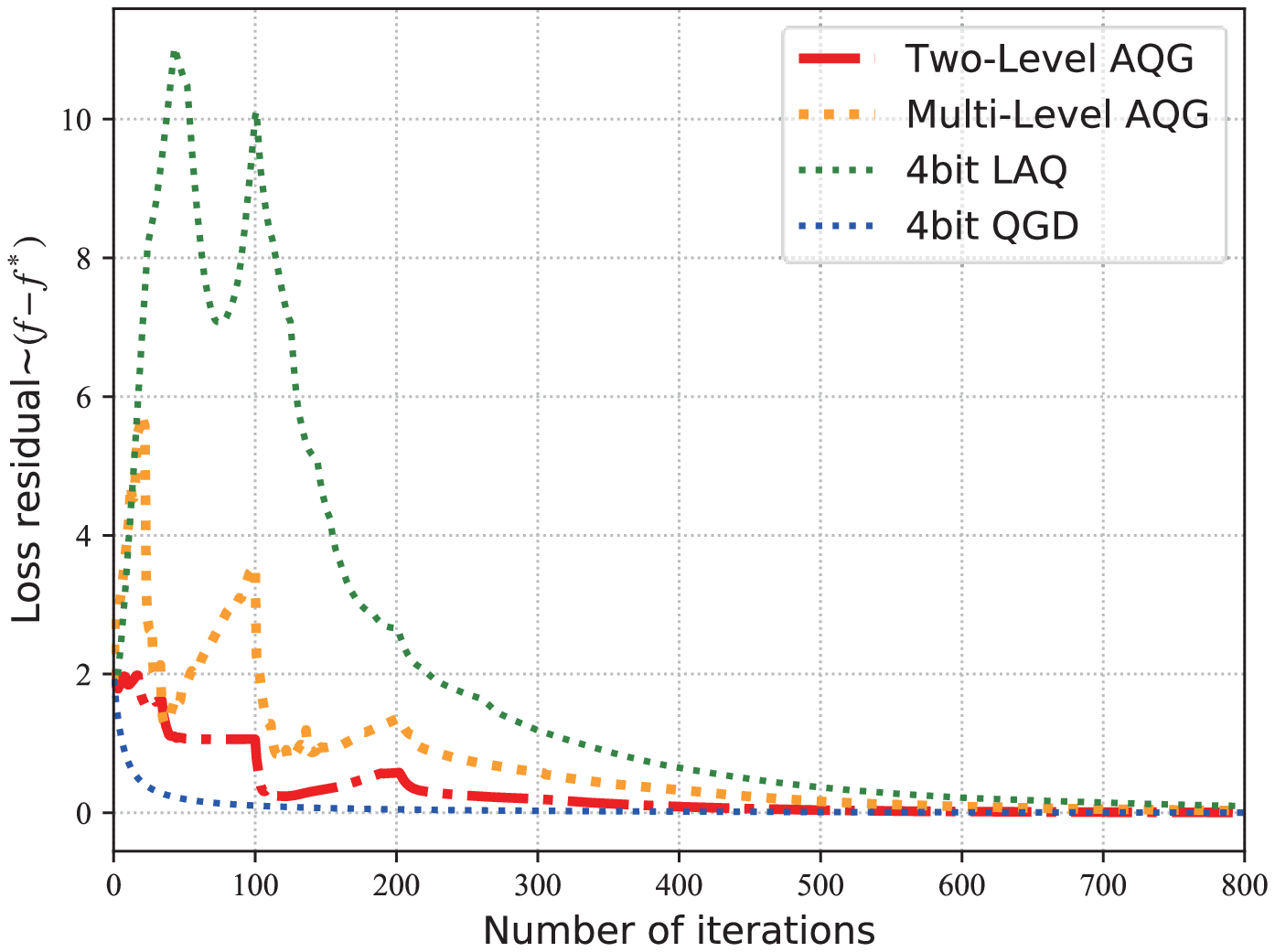}}
	\end{minipage} 
	\begin{minipage}[b]{.325\linewidth}
		\subfloat[][Loss v.s. communication round]{
			\label{4b}
			\includegraphics[width=1\linewidth]{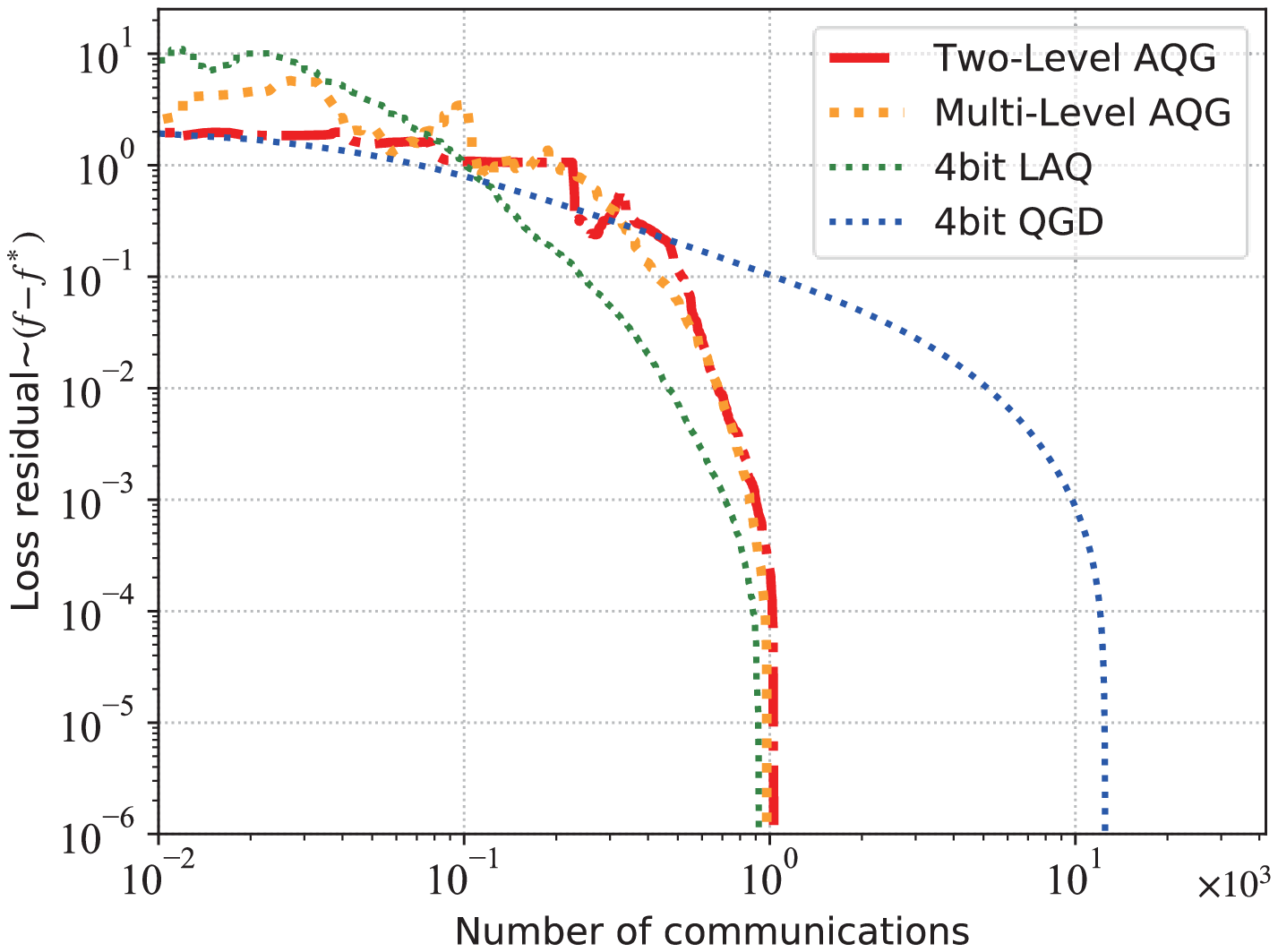}}
	\end{minipage} 
	\begin{minipage}[b]{.325\linewidth}
		\subfloat[][Loss v.s. bit]{
			\label{4c}
			\includegraphics[width=1\linewidth]{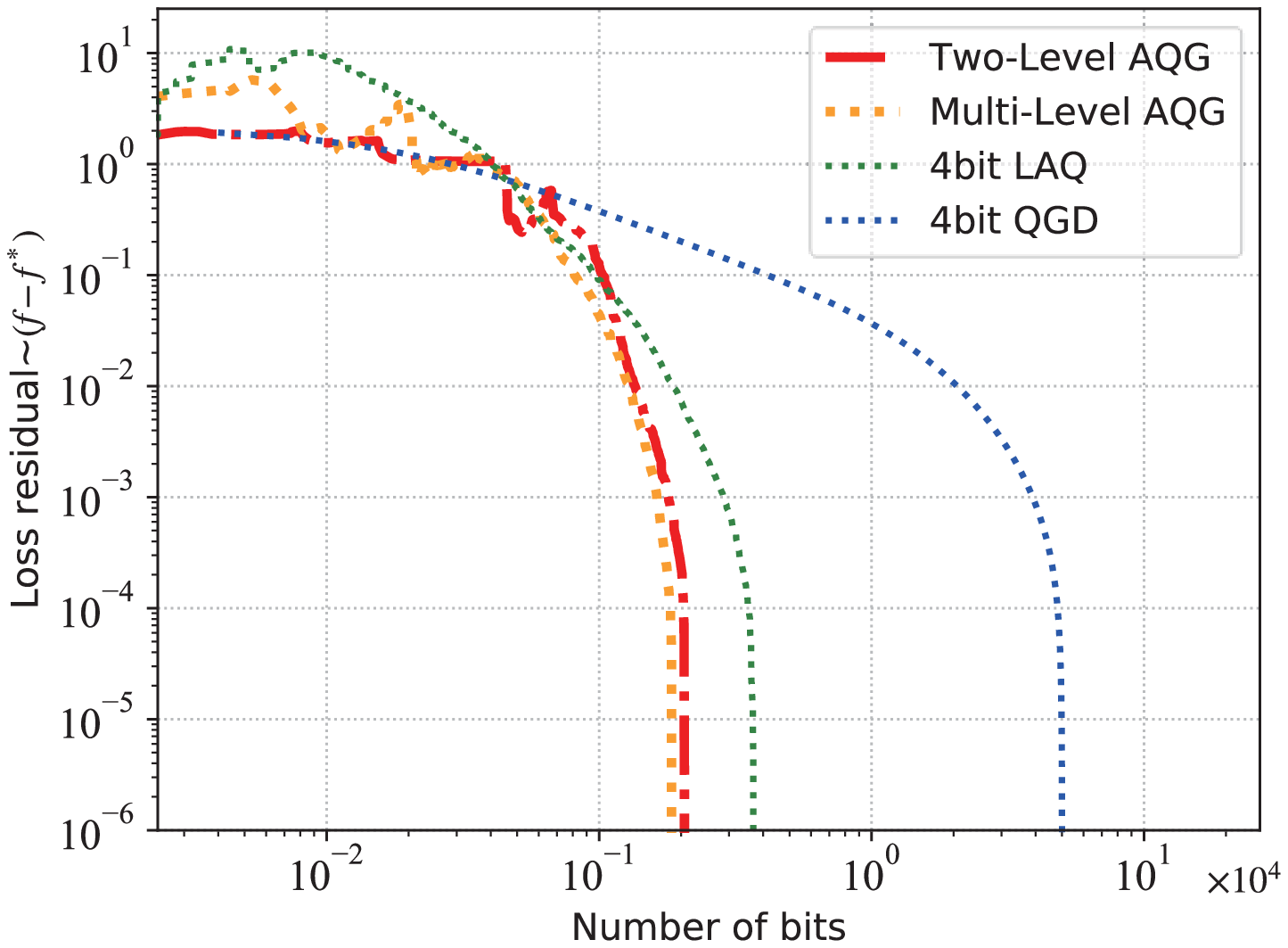}}
	\end{minipage} 
	\caption{Convergence of loss function with neural network and non-IID data distribution}
	\label{fig:network NonIID}
\end{figure}

\subsection{Performance of AQG with non-IID Data Distribution}

Fig.~\ref{fig:logistic NonIID} and Fig.~\ref{fig:network NonIID} verify that AQG works well with heterogeneous data distribution. Both variants of AQG manage to reduce the number of transmitted bits compared against other alternatives in both strongly convex and non-convex optimization. Meanwhile, it is obvious that experiments in non-IID data distribution benefit more with AQG compared against IID data distribution. The results are consistent with our expectation, since the idea of AQG is to utilize the inherent heterogeneousness of local optimization objectives.

\begin{figure}[htbp]
	\begin{minipage}[b]{.325\linewidth}
		\centering
		\subfloat[][Loss v.s. iteration]{
			\label{5a}
			\includegraphics[width=1\linewidth]{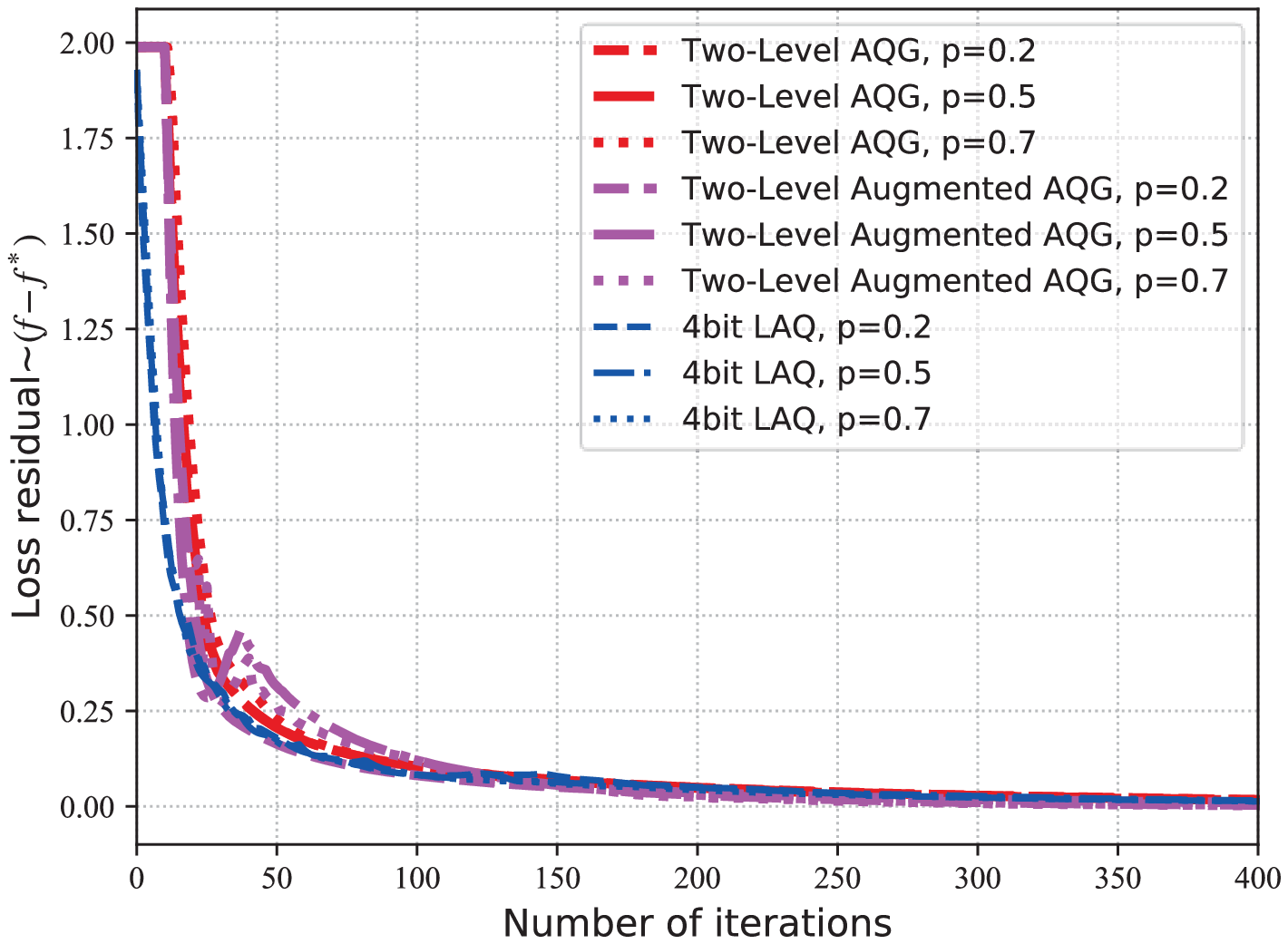}}
	\end{minipage} 
	\begin{minipage}[b]{.325\linewidth}
		\subfloat[][Loss v.s. communication round]{
			\label{5b}
			\includegraphics[width=1\linewidth]{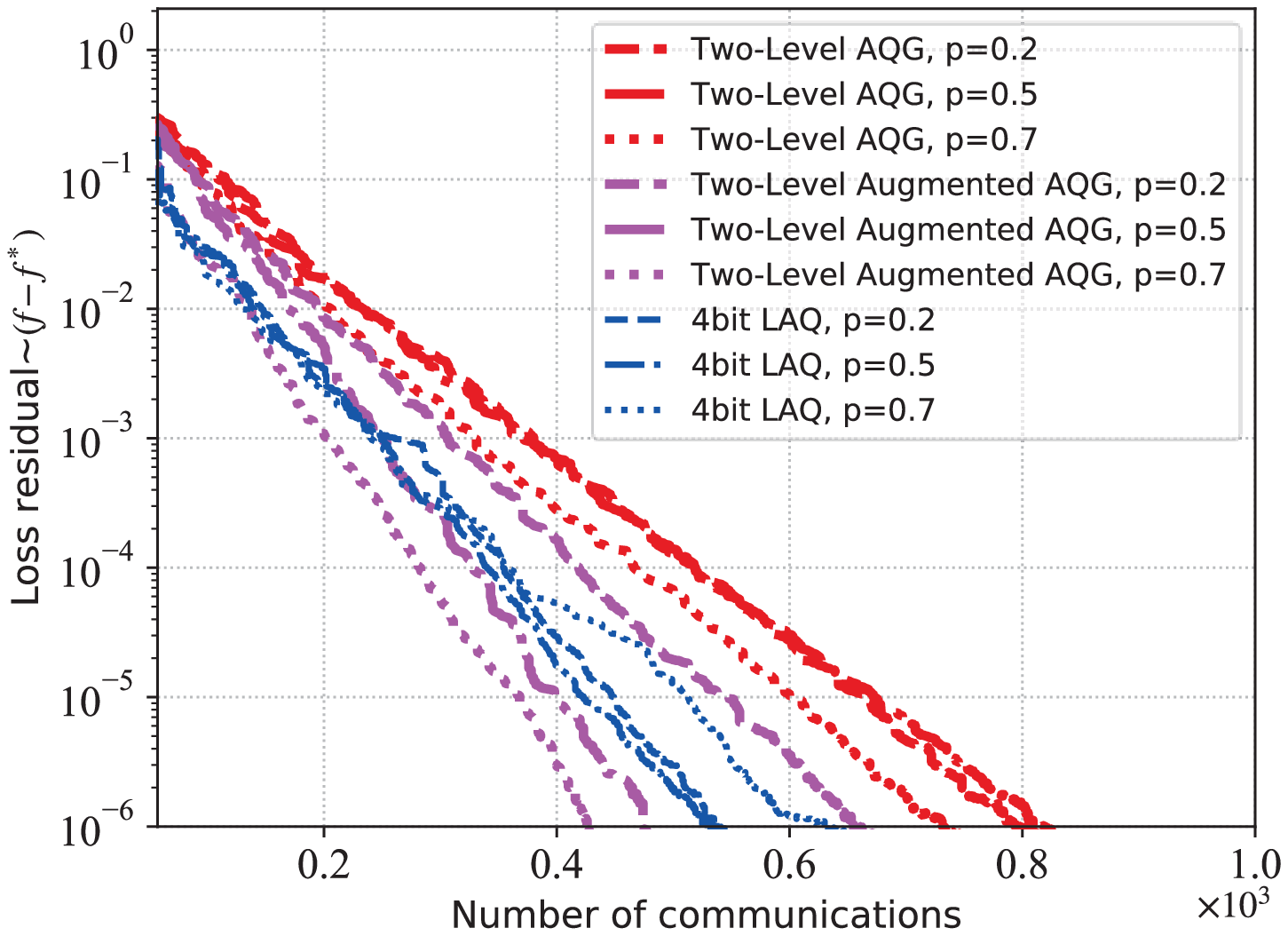}}
	\end{minipage} 
	\begin{minipage}[b]{.325\linewidth}
		\subfloat[][Loss v.s. bit]{
			\label{5c}
			\includegraphics[width=1\linewidth]{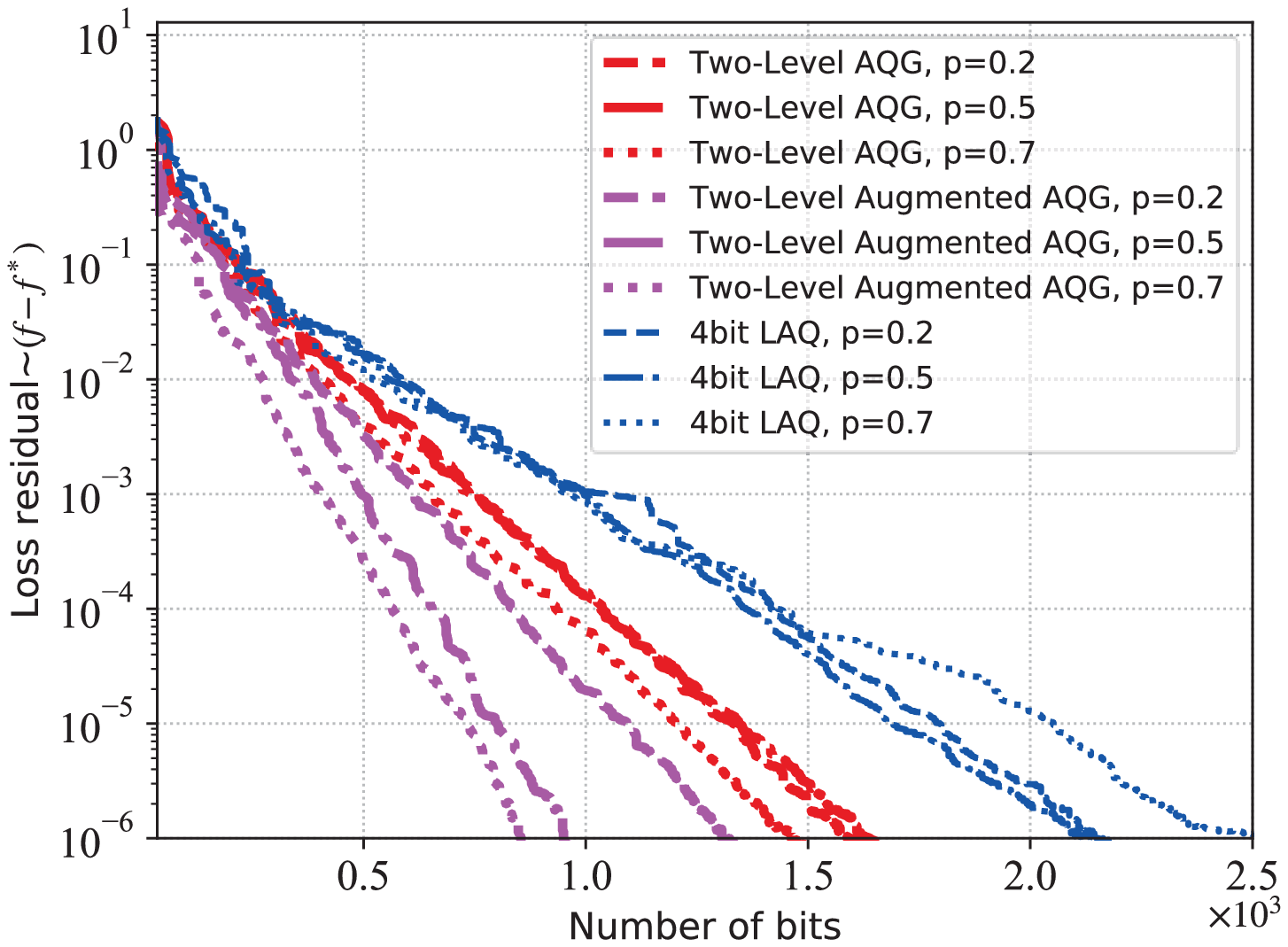}}
	\end{minipage} 
	\caption{Convergence of loss function with neural network \textbf{(p=0.2, 0.5 and 0.7)}.}
	\label{fig:305070}
\end{figure}

\begin{figure}[htbp]
	\begin{minipage}[b]{.325\linewidth}
		\centering
		\subfloat[][Loss v.s. iteration]{
			\label{6a}
			\includegraphics[width=1\linewidth]{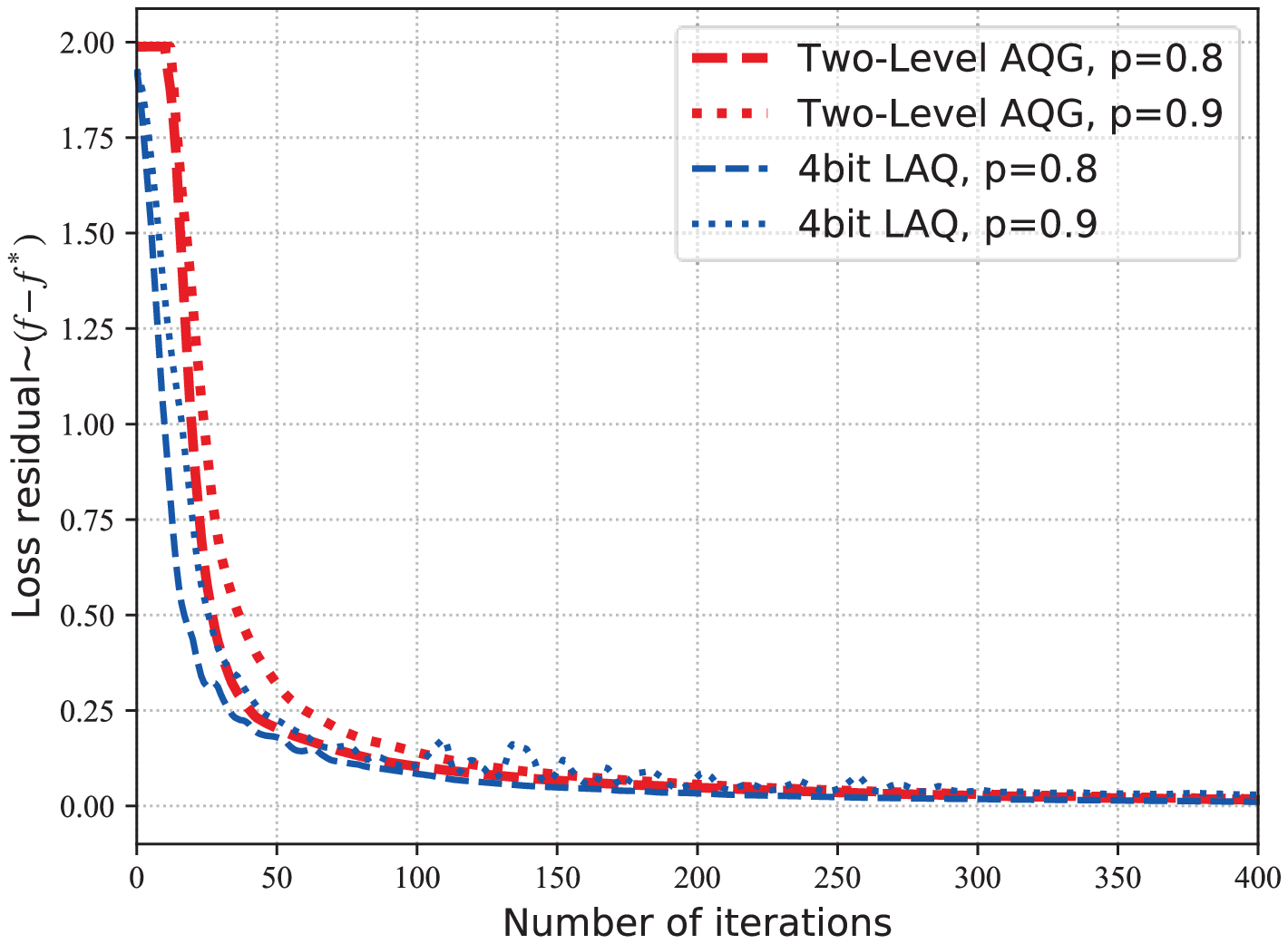}}
	\end{minipage} 
	\begin{minipage}[b]{.325\linewidth}
		\subfloat[][Loss v.s. communication round]{
			\label{6b}
			\includegraphics[width=1\linewidth]{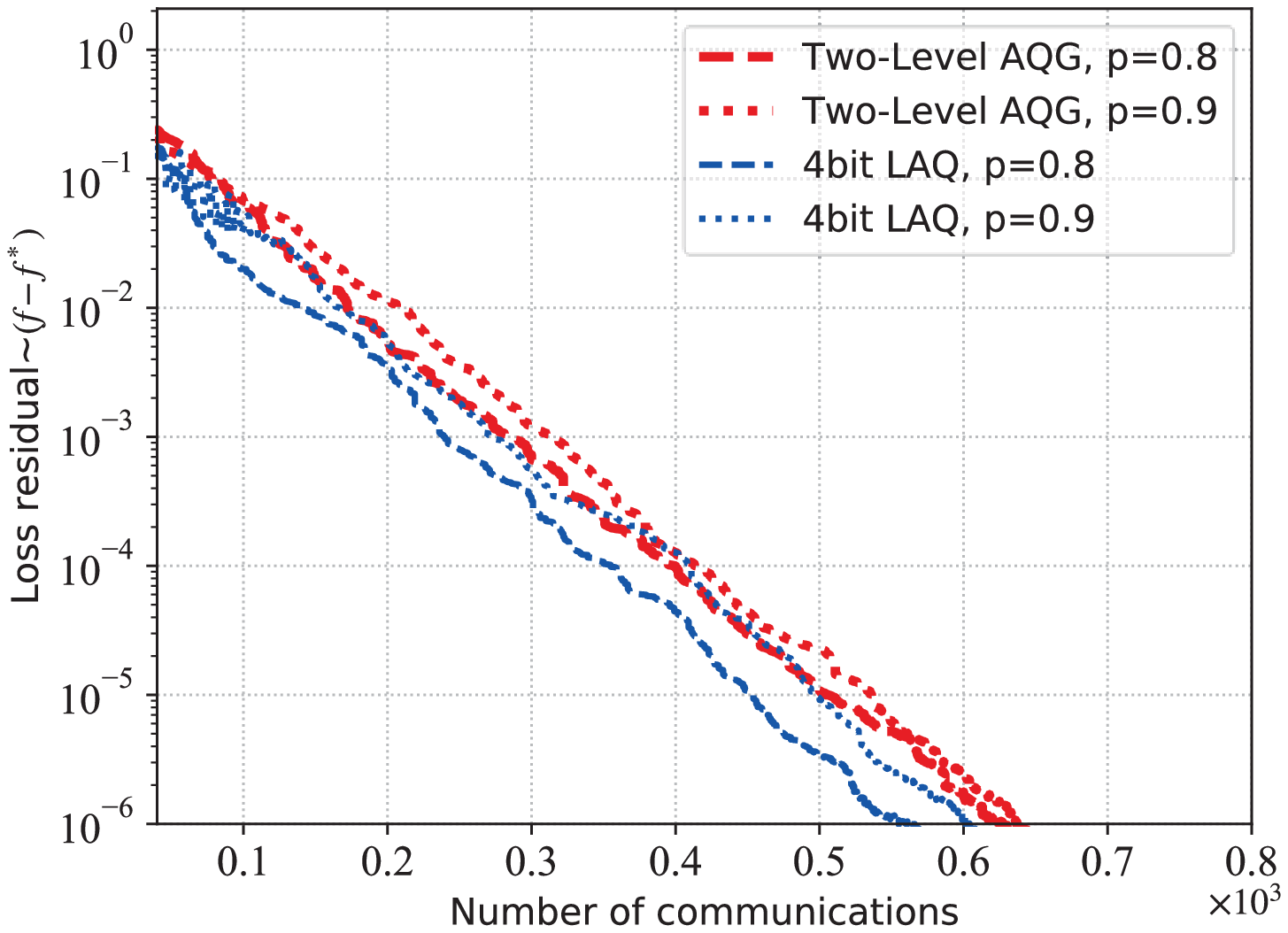}}
	\end{minipage} 
	\begin{minipage}[b]{.325\linewidth}
		\subfloat[][Loss v.s. bit]{
			\label{6c}
			\includegraphics[width=1\linewidth]{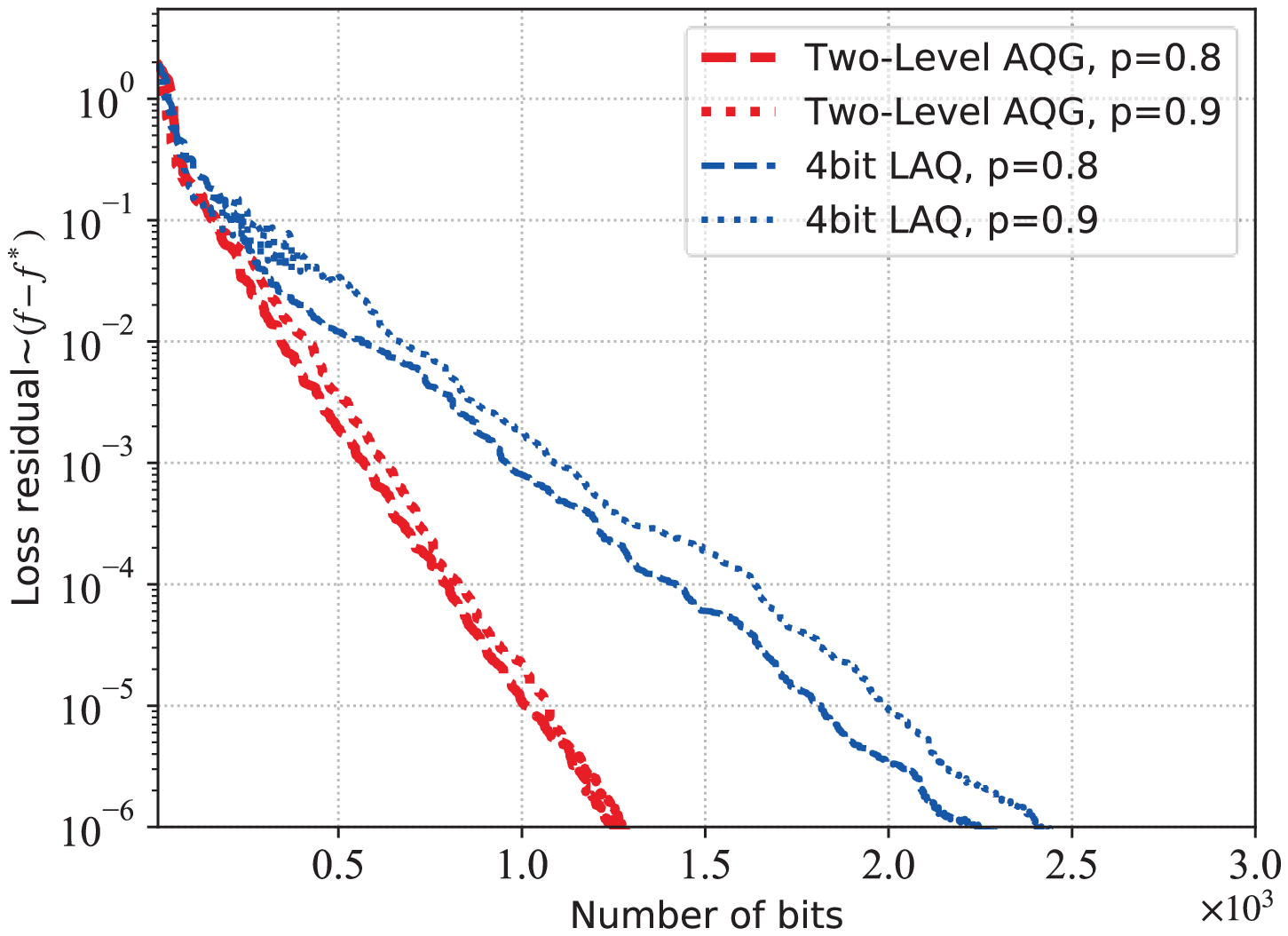}}
	\end{minipage} 
	\caption{Convergence of loss function with neural network \textbf{(p=0.8 and 0.9)}.}
	\label{fig:1020}
\end{figure}

\subsection{Performance of AQG with client dropouts}
In this part, we particularly focus on the setting of wireless network with mobile devices, where computation and communication are both extremely expensive, and client dropouts are frequent. Given these constraints, the Two-level AQG is applied in experiments with client dropouts as an adaptive solution for both communication and computation efficiency. Fig.~\ref{fig:305070} shows the performance of AQG with client dropping rate $p$ as 0.2, 0.5 and 0.7. Experiment results demonstrate that both AQG and Augmented AQG require fewer transmission bits compared against LAQ. Meanwhile, Augmented AQG has a stronger ability to reduce transmission bits with the presence of such moderate client dropouts.

Fig.~\ref{fig:1020} shows the performance of AQG with client dropping rate $p$ as 0.8 and 0.9. Experiments show that AQG manages to achieve stable convergence with ideal rates, and at the same time significantly reduces transmission bits even when there are only about 10$\%$ clients participating in gradient computation at each iteration. However, we notice that the augmented version of AQG fails to converge with a dropping rate higher than 0.8. It may be because when the dropping rate is too high, the unbiased estimation in Augmented AQG no longer remains accurate and even induces more noise into the training. Thus, the Augmented AQG is recommended to be applied in FL systems where the client dropping scale is moderate. Given the fact that the clients dropping rate is not likely to be so high in most practical systems, the augmented adaptive quantized gradient-based method is sufficient to address the dropping problem faced by FL.

\section{Conclusion}
\label{sec:conclusion}
This paper focuses on communication efficiency and the client dropout issue in FL, and proposes AQG which not only adaptively adjusts the quantization level depending on local gradient’s update before transmission, but also appropriately amplifies transmitted gradients to limit the dropout noise. For communication efficiency, the key idea is to quantize less informative gradient with less amount of bits, and vice versa. Since AQG fully utilize the heterogeneousness of local data distribution to reduce unnecessary transmission, it achieves a larger transmission reduction with non-IID data distribution as expected. Compared against existing popular methods, AQG leads to $25\%$-$50\%$ of transmission reduction while keeping the desired convergence properties, and shows robustness to large-scale client dropouts with a dropping rate up to $90\%$. Meanwhile, the Augmented AQG brings extra transmission reduction with moderate-scale client dropouts commonly seen in practical scenarios, which indicates gradient amplification’s effectiveness in suppressing the noise introduced by client dropouts. 

Due to the aforementioned superiorities, AQG can be used jointly with some other communication efficient methods for FL architectures, such as gradient sparsification\cite{Sparsification5}, client selection based on local resources\cite{Clientselection1, Clientselection2, Clientselection3} and adaptively distributing subnetworks for heterogeneous clients \cite{diao2020heterofl, bouacida2020adaptive}. Such superiorities and flexibility endow great potentials for the proposed FL framework with AQG. Future works include deploying AQG jointly with such techniques in practical FL systems.

\begin{acks}
	This work is supported in part by \grantsponsor{}{Tsinghua-Foshan Innovation Special Fund (TFISF)}{} under Grant
	No.\grantnum{}{2020THFS0109}
	and \grantsponsor{}{Guangdong Basic and Applied Basic Research Foundation}{} under Grant
	No.\grantnum{}{2020A1515110887}.
\end{acks}

\bibliographystyle{ACM-Reference-Format}
\bibliography{MyRef}

\clearpage
\allowdisplaybreaks[4]
\appendix

\section{Mathematical Proof}

\subsection{Proof of Lemma 1}

In AQG:

\begin{align}
&\sum_{m=1}^{M}Q_{\hat{b}_{m}^{k}}(\hat{\boldsymbol{g}}_{m}^{k}) = \sum_{b=1}^{b_{max}}\sum_{m\in \mathbb{M}_{b}^{k}}Q_{\hat{b}_{m}^{k}}(\hat{\boldsymbol{g}}_{m}^{k})+\sum_{m\in \mathbb{M}_{0}^{k}}Q_{\hat{b}_{m}^{k-1}}(\hat{\boldsymbol{g}}_{m}^{k-1})  \nonumber \\ &=\sum_{m=1}^{M}Q_{b_{max}}(\hat{\boldsymbol{g}}_{m}^{k})+\sum_{b=1}^{b_{max}}\sum_{m\in \mathbb{M}_{b}^{k}}[Q_{\hat{b}_{m}^{k}}(\hat{\boldsymbol{g}}_{m}^{k})-Q_{b_{max}}(\hat{\boldsymbol{g}}_{m}^{k})]  
+ \sum_{m\in \mathbb{M}_{0}^{k}}[Q_{\hat{b}_{m}^{k-1}}(\hat{\boldsymbol{g}}_{m}^{k-1})-Q_{b_{max}}(\hat{\boldsymbol{g}}_{m}^{k})]
\end{align}

From the update rule of AQG:
\begin{equation}
\boldsymbol{\theta} ^{k+1} - \boldsymbol{\theta} ^{k} = -\alpha \sum_{m=1}^{M}Q_{\hat{b}_{m}^{k}}(\hat{\boldsymbol{g}}_{m}^{k})
\end{equation}

From the definition of quantization error:
\begin{equation}
\sum_{m=1}^{M}Q_{b_{max}}(\hat{\boldsymbol{g}}_{m}^{k}) = \nabla f(\boldsymbol{\theta} ^{k})-\sum_{m=1}^{M}\varepsilon_{b_{max}}(\hat{\boldsymbol{g}}_{m}^{k})
\label{eq:app1}
\end{equation}

With inequality $\left \langle \boldsymbol{a}, \boldsymbol{b} \right \rangle\leq \frac{1}{2}\rho\left \| \boldsymbol{a} \right \|_{2}^{2} + \frac{1}{2\rho}\left \| \boldsymbol{b} \right \|_{2}^{2}$ and \eqref{eq:app1}:
\begin{align}
&-\alpha \left \langle \nabla f(\boldsymbol{\theta} ^{k}),\sum_{m=1}^{M}Q_{b_{max}}(\hat{\boldsymbol{g}}_{m}^{k}) \right \rangle = -\alpha \left \langle \nabla f(\boldsymbol{\theta} ^{k}),\nabla f(\boldsymbol{\theta} ^{k})-\sum_{m=1}^{M}\varepsilon_{b_{max}}(\hat{\boldsymbol{g}}_{m}^{k}) \right \rangle \nonumber \\
&= -\alpha \left \| \nabla f(\boldsymbol{\theta} ^{k}) \right \|_{2}^{2}+\alpha \left \langle \nabla f(\boldsymbol{\theta} ^{k}),\sum_{m=1}^{M}\varepsilon_{b_{max}}(\hat{\boldsymbol{g}}_{m}^{k}) \right \rangle \nonumber \\
& \leq -\alpha \left \| \nabla f(\boldsymbol{\theta} ^{k}) \right \|_{2}^{2} + \frac{\alpha \rho_{1}}{2}\left \| \nabla f(\boldsymbol{\theta} ^{k}) \right \|_{2}^{2} + \frac{\alpha}{2\rho_{1}}\left \| \sum_{m=1}^{M}\varepsilon_{b_{max}}(\hat{\boldsymbol{g}}_{m}^{k}) \right \|_{2}^{2}
\end{align}

Under Assumption 1:
\begin{align}
&f(\boldsymbol{\theta} ^{k+1})-f(\boldsymbol{\theta} ^{k})\leq \left \langle \nabla f(\boldsymbol{\theta} ^{k}), \boldsymbol{\theta} ^{k+1}-\boldsymbol{\theta} ^{k} \right \rangle + \frac{L}{2}\left \| \boldsymbol{\theta} ^{k+1}-\boldsymbol{\theta} ^{k} \right \|_{2}^{2} \nonumber \\
&= \left \langle \nabla f(\boldsymbol{\theta} ^{k}), -\alpha \sum_{m=1}^{M}Q_{\hat{b}_{m}^{k}}(\hat{\boldsymbol{g}}_{m}^{k}) \right \rangle + \frac{L}{2}\left \| \boldsymbol{\theta} ^{k+1}-\boldsymbol{\theta} ^{k} \right \|_{2}^{2}  \nonumber \\
&=  \left \langle \nabla f(\boldsymbol{\theta} ^{k}), -\alpha  \sum_{m=1}^{M}Q_{b_{max}}(\hat{\boldsymbol{g}}_{m}^{k}) \right \rangle +\frac{L}{2}\left \| \boldsymbol{\theta} ^{k+1}-\boldsymbol{\theta} ^{k} \right \|_{2}^{2} \nonumber \\
&+  \left \langle \nabla f(\boldsymbol{\theta} ^{k}), -\alpha \left\{ \sum_{b=1}^{b_{max}}\sum_{m\in \mathbb{M}_{b}^{k}}[Q_{\hat{b}_{m}^{k}}(\hat{\boldsymbol{g}}_{m}^{k})-Q_{b_{max}}(\hat{\boldsymbol{g}}_{m}^{k})] 
+ \sum_{m\in \mathbb{M}_{0}^{k}}[Q_{\hat{b}_{m}^{k-1}}(\hat{\boldsymbol{g}}_{m}^{k-1})-Q_{b_{max}}(\hat{\boldsymbol{g}}_{m}^{k})]
\right\} \right \rangle \nonumber \\ 
& \leq \left \langle \nabla f(\boldsymbol{\theta} ^{k}), -\alpha  \sum_{m=1}^{M}Q_{b_{max}}(\hat{\boldsymbol{g}}_{m}^{k}) \right \rangle +\frac{L}{2}\left \| \boldsymbol{\theta} ^{k+1}-\boldsymbol{\theta} ^{k} \right \|_{2}^{2} \nonumber +\frac{\alpha}{2}\left \| \nabla f(\boldsymbol{\theta} ^{k}) \right \|_{2}^{2} \nonumber \\
& + \frac{\alpha}{2}\left \| \sum_{b=1}^{b_{max}}\sum_{m\in \mathbb{M}_{b}^{k}}[Q_{\hat{b}_{m}^{k}}(\hat{\boldsymbol{g}}_{m}^{k})-Q_{b_{max}}(\hat{\boldsymbol{g}}_{m}^{k})]  
+ \sum_{m\in \mathbb{M}_{0}^{k}}[Q_{\hat{b}_{m}^{k-1}}(\hat{\boldsymbol{g}}_{m}^{k-1})-Q_{b_{max}}(\hat{\boldsymbol{g}}_{m}^{k})] \right \|_{2}^{2} 
\end{align}

The Lyapunov function of AQG is defined as:
\begin{equation}
\mathbb{V}(\boldsymbol{\theta} ^{k})=f(\boldsymbol{\theta} ^{k})-f(\boldsymbol{\theta} ^{*})+\sum_{d=1}^{D}\sum_{j=d}^{D}\frac{\xi  _{j}}{\alpha }\left \| \boldsymbol{\theta} ^{k+1-d}- \boldsymbol{\theta} ^{k-d} \right \|_{2}^{2} 
\end{equation}

Let $\beta _{d}=\frac{1}{\alpha }\sum_{j=d}^{D}\xi _{j}, \forall d \in \{1,...,D\}$, then:
\begin{equation}
\mathbb{V}(\boldsymbol{\theta} ^{k})=f(\boldsymbol{\theta} ^{k})-f(\boldsymbol{\theta} ^{*})+\sum_{d=1}^{D}\beta_{d}\left \| \boldsymbol{\theta} ^{k+1-d}- \boldsymbol{\theta} ^{k-d} \right \|_{2}^{2} 
\end{equation}

Thus,
\begin{align}
&\mathbb{V}(\boldsymbol{\theta} ^{k+1})-\mathbb{V}(\boldsymbol{\theta} ^{k}) =  f(\boldsymbol{\theta} ^{k+1})-f(\boldsymbol{\theta} ^{k})+\sum_{d=1}^{D}\beta_{d}\left \| \boldsymbol{\theta} ^{k+1-(d-1)}- \boldsymbol{\theta} ^{k-(d-1)} \right \|_{2}^{2}-\sum_{d=1}^{D}\beta_{d}\left \| \boldsymbol{\theta} ^{k+1-d}- \boldsymbol{\theta} ^{k-d} \right \|_{2}^{2} \nonumber \\
&= f(\boldsymbol{\theta} ^{k+1})-f(\boldsymbol{\theta} ^{k})+\beta_{1}\left \| \boldsymbol{\theta} ^{k+1}- \boldsymbol{\theta} ^{k} \right \|_{2}^{2}+\sum_{d=1}^{D-1}(\beta_{d+1}-\beta_{d})\left \| \boldsymbol{\theta} ^{k+1-d}- \boldsymbol{\theta} ^{k-d} \right \|_{2}^{2}-\beta_{D}\left \| \boldsymbol{\theta} ^{k+1-D}- \boldsymbol{\theta} ^{k-D} \right \|_{2}^{2} \nonumber \\
& \leq -\alpha \left \langle \nabla f(\boldsymbol{\theta} ^{k}),\sum_{m=1}^{M}Q_{b_{max}}(\hat{\boldsymbol{g}}_{m}^{k}) \right \rangle + \frac{\alpha}{2}\left \| \nabla f(\boldsymbol{\theta} ^{k}) \right \|_{2}^{2} \nonumber \\
& + \frac{\alpha}{2}\left \| \sum_{b=1}^{b_{max}}\sum_{m\in \mathbb{M}_{b}^{k}}[Q_{\hat{b}_{m}^{k}}(\hat{\boldsymbol{g}}_{m}^{k})-Q_{b_{max}}(\hat{\boldsymbol{g}}_{m}^{k})]  
+ \sum_{m\in \mathbb{M}_{0}^{k}}[Q_{\hat{b}_{m}^{k-1}}(\hat{\boldsymbol{g}}_{m}^{k-1})-Q_{b_{max}}(\hat{\boldsymbol{g}}_{m}^{k})] \right \|_{2}^{2} \nonumber \\
&+(\frac{L}{2}+\beta_{1})\left \| \boldsymbol{\theta} ^{k+1}- \boldsymbol{\theta} ^{k} \right \|_{2}^{2}+\sum_{d=1}^{D-1}(\beta_{d+1}-\beta_{d})\left \| \boldsymbol{\theta} ^{k+1-d}- \boldsymbol{\theta} ^{k-d} \right \|_{2}^{2}-\beta_{D}\left \| \boldsymbol{\theta} ^{k+1-D}- \boldsymbol{\theta} ^{k-D} \right \|_{2}^{2} \\
& = -\alpha \left \langle \nabla f(\boldsymbol{\theta} ^{k}),\sum_{m=1}^{M}Q_{b_{max}}(\hat{\boldsymbol{g}}_{m}^{k}) \right \rangle + \frac{\alpha}{2}\left \| \nabla f(\boldsymbol{\theta} ^{k}) \right \|_{2}^{2} +\sum_{d=1}^{D-1}(\beta_{d+1}-\beta_{d})\left \| \boldsymbol{\theta} ^{k+1-d}- \boldsymbol{\theta} ^{k-d} \right \|_{2}^{2}-\beta_{D}\left \| \boldsymbol{\theta} ^{k+1-D}- \boldsymbol{\theta} ^{k-D} \right \|_{2}^{2}\nonumber \\
&+(\frac{L}{2}+\beta_{1})\left \| \alpha \left\{ \sum_{m=1}^{M}Q_{b_{max}}(\hat{\boldsymbol{g}}_{m}^{k})+\sum_{b=1}^{b_{max}}\sum_{m\in \mathbb{M}_{b}^{k}}[Q_{\hat{b}_{m}^{k}}(\hat{\boldsymbol{g}}_{m}^{k})-Q_{b_{max}}(\hat{\boldsymbol{g}}_{m}^{k})]  
+ \sum_{m\in \mathbb{M}_{0}^{k}}[Q_{\hat{b}_{m}^{k-1}}(\hat{\boldsymbol{g}}_{m}^{k-1})-Q_{b_{max}}(\hat{\boldsymbol{g}}_{m}^{k})]\right\} \right \|_{2}^{2} \nonumber \\
& + \frac{\alpha}{2}\left \| \sum_{b=1}^{b_{max}}\sum_{m\in \mathbb{M}_{b}^{k}}[Q_{\hat{b}_{m}^{k}}(\hat{\boldsymbol{g}}_{m}^{k})-Q_{b_{max}}(\hat{\boldsymbol{g}}_{m}^{k})]  
+ \sum_{m\in \mathbb{M}_{0}^{k}}[Q_{\hat{b}_{m}^{k-1}}(\hat{\boldsymbol{g}}_{m}^{k-1})-Q_{b_{max}}(\hat{\boldsymbol{g}}_{m}^{k})] \right \|_{2}^{2} 
\end{align}

From Young's Equality $\left \| \textbf{a}+\mathbf{b} \right \|_{2}^{2}\leq (1+\rho)\left \| \textbf{a} \right \|_{2}^{2} + (1+\rho^{-1})\left \| \textbf{b} \right \|_{2}^{2}$, there is:
\begin{align}
&(\frac{L}{2}+\beta_{1})\left \| \alpha \left\{ \sum_{m=1}^{M}Q_{b_{max}}(\hat{\boldsymbol{g}}_{m}^{k})+\sum_{b=1}^{b_{max}}\sum_{m\in \mathbb{M}_{b}^{k}}[Q_{\hat{b}_{m}^{k}}(\hat{\boldsymbol{g}}_{m}^{k})-Q_{b_{max}}(\hat{\boldsymbol{g}}_{m}^{k})]  
+ \sum_{m\in \mathbb{M}_{0}^{k}}[Q_{\hat{b}_{m}^{k-1}}(\hat{\boldsymbol{g}}_{m}^{k-1})-Q_{b_{max}}(\hat{\boldsymbol{g}}_{m}^{k})]\right\} \right \|_{2}^{2} \nonumber \\
& \leq (\frac{L}{2}+\beta_{1})(1+\rho _{2}^{-1}) \alpha^{2} \left \| \sum_{b=1}^{b_{max}}\sum_{m\in \mathbb{M}_{b}^{k}}[Q_{\hat{b}_{m}^{k}}(\hat{\boldsymbol{g}}_{m}^{k})-Q_{b_{max}}(\hat{\boldsymbol{g}}_{m}^{k})]  
+ \sum_{m\in \mathbb{M}_{0}^{k}}[Q_{\hat{b}_{m}^{k-1}}(\hat{\boldsymbol{g}}_{m}^{k-1})-Q_{b_{max}}(\hat{\boldsymbol{g}}_{m}^{k})] \right \|_{2}^{2} \nonumber \\
& + (\frac{L}{2}+\beta_{1})(1+\rho _{2}) \alpha^{2} \left \| \sum_{m=1}^{M}Q_{b_{max}}(\hat{\boldsymbol{g}}_{m}^{k}) \right \|_{2}^{2} 
\label{eq:25}
\end{align}

From $\left \| \sum_{i=1}^{n}\boldsymbol{a}_{i} \right \|_{2}^{2}\leq n \sum_{i=1}^{n}\left \| \boldsymbol{a}_{i} \right \|_{2}^{2}$, there is:
\begin{align}
&\left \| \sum_{b=1}^{b_{max}}\sum_{m\in \mathbb{M}_{b}^{k}}[Q_{\hat{b}_{m}^{k}}(\hat{\boldsymbol{g}}_{m}^{k})-Q_{b_{max}}(\hat{\boldsymbol{g}}_{m}^{k})]  
+ \sum_{m\in \mathbb{M}_{0}^{k}}[Q_{\hat{b}_{m}^{k-1}}(\hat{\boldsymbol{g}}_{m}^{k-1})-Q_{b_{max}}(\hat{\boldsymbol{g}}_{m}^{k})] \right \|_{2}^{2} \nonumber  \\
&\leq M \sum_{b=1}^{b_{max}}\sum_{m\in \mathbb{M}_{b}^{k}} \left \| Q_{\hat{b}_{m}^{k}}(\hat{\boldsymbol{g}}_{m}^{k})-Q_{b_{max}}(\hat{\boldsymbol{g}}_{m}^{k}) \right \|_{2}^{2} + M\sum_{m\in \mathbb{M}_{0}^{k}}\left \| Q_{\hat{b}_{m}^{k-1}}(\hat{\boldsymbol{g}}_{m}^{k-1})-Q_{b_{max}}(\hat{\boldsymbol{g}}_{m}^{k}) \right \|_{2}^{2} \nonumber \\
& = 2 M \sum_{b=1}^{b_{max}}\sum_{m\in \mathbb{M}_{b}^{k}} \left \| \varepsilon _{\hat{b}_{m}^{k}}(\hat{\boldsymbol{g}}_{m}^{k}) \right \|_{2}^{2} + 2 M \sum_{b=1}^{b_{max}}\sum_{m\in \mathbb{M}_{b}^{k}} \left \| \varepsilon _{b_{max}}(\hat{\boldsymbol{g}}_{m}^{k}) \right \|_{2}^{2} +M\sum_{m\in \mathbb{M}_{0}^{k}}\left \| Q_{\hat{b}_{m}^{k-1}}(\hat{\boldsymbol{g}}_{m}^{k-1})-Q_{b_{max}}(\hat{\boldsymbol{g}}_{m}^{k}) \right \|_{2}^{2}
\label{eq:26}
\end{align}

With \eqref{eq:25} and \eqref{eq:26}:
\begin{align}
&\mathbb{V}(\boldsymbol{\theta} ^{k+1})-\mathbb{V}(\boldsymbol{\theta} ^{k}) \leq -\alpha \left \langle \nabla f(\boldsymbol{\theta} ^{k}),\sum_{m=1}^{M}Q_{b_{max}}(\hat{\boldsymbol{g}}_{m}^{k}) \right \rangle + \frac{\alpha}{2}\left \| \nabla f(\boldsymbol{\theta} ^{k}) \right \|_{2}^{2} \nonumber \\
& + (\frac{L}{2}+\beta_{1})(1+\rho _{2}) \alpha^{2} \left \| \sum_{m=1}^{M}Q_{b_{max}}(\hat{\boldsymbol{g}}_{m}^{k}) \right \|_{2}^{2} +\sum_{d=1}^{D-1}(\beta_{d+1}-\beta_{d})\left \| \boldsymbol{\theta} ^{k+1-d}- \boldsymbol{\theta} ^{k-d} \right \|_{2}^{2}-\beta_{D}\left \| \boldsymbol{\theta} ^{k+1-D}- \boldsymbol{\theta} ^{k-D} \right \|_{2}^{2}\nonumber \\
& + [\frac{\alpha}{2}+(\frac{L}{2}+\beta_{1})(1+\rho _{2}^{-1}) \alpha^{2}] \left \| \sum_{b=1}^{b_{max}}\sum_{m\in \mathbb{M}_{b}^{k}}[Q_{\hat{b}_{m}^{k}}(\hat{\boldsymbol{g}}_{m}^{k})-Q_{b_{max}}(\hat{\boldsymbol{g}}_{m}^{k})]  
+ \sum_{m\in \mathbb{M}_{0}^{k}}[Q_{\hat{b}_{m}^{k-1}}(\hat{\boldsymbol{g}}_{m}^{k-1})-Q_{b_{max}}(\hat{\boldsymbol{g}}_{m}^{k})] \right \|_{2}^{2} \nonumber \\ 
&\leq -\alpha \left \langle \nabla f(\boldsymbol{\theta} ^{k}),\sum_{m=1}^{M}Q_{b_{max}}(\hat{\boldsymbol{g}}_{m}^{k}) \right \rangle + \frac{\alpha}{2}\left \| \nabla f(\boldsymbol{\theta} ^{k}) \right \|_{2}^{2}  + (\frac{L}{2}+\beta_{1})(1+\rho _{2}) \alpha^{2} \left \| \sum_{m=1}^{M}Q_{b_{max}}(\hat{\boldsymbol{g}}_{m}^{k}) \right \|_{2}^{2} \nonumber \\
&+\sum_{d=1}^{D-1}(\beta_{d+1}-\beta_{d})\left \| \boldsymbol{\theta} ^{k+1-d}- \boldsymbol{\theta} ^{k-d} \right \|_{2}^{2}-\beta_{D}\left \| \boldsymbol{\theta} ^{k+1-D}- \boldsymbol{\theta} ^{k-D} \right \|_{2}^{2} \nonumber \\
& +  2[\frac{\alpha}{2}+(\frac{L}{2}+\beta_{1})(1+\rho _{2}^{-1}) \alpha^{2}]M( \sum_{b=1}^{b_{max}}\sum_{m\in \mathbb{M}_{b}^{k}} \left \| \varepsilon _{\hat{b}_{m}^{k}}(\hat{\boldsymbol{g}}_{m}^{k}) \right \|_{2}^{2} +  \sum_{b=1}^{b_{max}}\sum_{m\in \mathbb{M}_{b}^{k}} \left \| \varepsilon _{b_{max}}(\hat{\boldsymbol{g}}_{m}^{k}) \right \|_{2}^{2}) \nonumber \\ 
& +  [\frac{\alpha}{2}+(\frac{L}{2}+\beta_{1})(1+\rho _{2}^{-1}) \alpha^{2}]M\sum_{m\in \mathbb{M}_{0}^{k}}\left \| Q_{\hat{b}_{m}^{k-1}}(\hat{\boldsymbol{g}}_{m}^{k-1})-Q_{b_{max}}(\hat{\boldsymbol{g}}_{m}^{k}) \right \|_{2}^{2} 
\end{align}

With the precision selection criterion \eqref{eq6}:
\begin{align}
&M\sum_{m\in \mathbb{M}_{0}^{k}} \left \| Q_{\hat{b}_{m}^{k-1}}(\hat{\boldsymbol{g}}_{m}^{k-1})-Q_{b_{max}}(\boldsymbol{g}_{m}^{k}) \right \|_{2}^{2}  \nonumber  \\
&\leq\frac{M^{2}}{\alpha ^{2}M^{2}}\sum_{d=1}^{D}\xi _{d}\left \| \boldsymbol{\theta} ^{k+1-d}-\boldsymbol{\theta} ^{k-d} \right \|_{2}^{2} +3M\sum_{m\in \mathbb{M}_{0}^{k}}(\left \| \varepsilon _{b_{max}}(\hat{\boldsymbol{g}}_{m}^{k-1})\right \|_{2}^{2}+\left \| \varepsilon _{b_{max}}({\boldsymbol{g}}_{m}^{k}) \right \|_{2}^{2}) 
\end{align}

Thus,
\begin{align}
&\mathbb{V}(\boldsymbol{\theta} ^{k+1})-\mathbb{V}(\boldsymbol{\theta} ^{k}) \nonumber \\
&\leq -\alpha \left \langle \nabla f(\boldsymbol{\theta} ^{k}),\sum_{m=1}^{M}Q_{b_{max}}(\hat{\boldsymbol{g}}_{m}^{k}) \right \rangle + \frac{\alpha}{2}\left \| \nabla f(\boldsymbol{\theta} ^{k}) \right \|_{2}^{2}  + (\frac{L}{2}+\beta_{1})(1+\rho _{2}) \alpha^{2} \left \| \sum_{m=1}^{M}Q_{b_{max}}(\hat{\boldsymbol{g}}_{m}^{k}) \right \|_{2}^{2} \nonumber \\
&+\sum_{d=1}^{D-1}(\beta_{d+1}-\beta_{d})\left \| \boldsymbol{\theta} ^{k+1-d}- \boldsymbol{\theta} ^{k-d} \right \|_{2}^{2}-\beta_{D}\left \| \boldsymbol{\theta} ^{k+1-D}- \boldsymbol{\theta} ^{k-D} \right \|_{2}^{2} \nonumber \\
& +  2[\frac{\alpha}{2}+(\frac{L}{2}+\beta_{1})(1+\rho _{2}^{-1}) \alpha^{2}]M( \sum_{b=1}^{b_{max}}\sum_{m\in \mathbb{M}_{b}^{k}} \left \| \varepsilon _{\hat{b}_{m}^{k}}(\hat{\boldsymbol{g}}_{m}^{k}) \right \|_{2}^{2} +  \sum_{b=1}^{b_{max}}\sum_{m\in \mathbb{M}_{b}^{k}} \left \| \varepsilon _{b_{max}}(\hat{\boldsymbol{g}}_{m}^{k}) \right \|_{2}^{2}) \nonumber \\ 
& +  [\frac{\alpha}{2}+(\frac{L}{2}+\beta_{1})(1+\rho _{2}^{-1}) \alpha^{2}]\frac{1}{\alpha ^{2}}\sum_{d=1}^{D}\xi _{d}\left \| \boldsymbol{\theta} ^{k+1-d}-\boldsymbol{\theta} ^{k-d} \right \|_{2}^{2}  \nonumber \\ 
& + 3[\frac{\alpha}{2}+(\frac{L}{2}+\beta_{1})(1+\rho _{2}^{-1}) \alpha^{2}]M\sum_{m\in \mathbb{M}_{0}^{k}}(\left \| \varepsilon _{b_{max}}(\hat{\boldsymbol{g}}_{m}^{k-1})\right \|_{2}^{2}+\left \| \varepsilon _{b_{max}}({\boldsymbol{g}}_{m}^{k}) \right \|_{2}^{2}) \\
&\leq (-\frac{\alpha}{2}+\frac{\alpha\rho_{1}}{2})\left \| \nabla f(\boldsymbol{\theta} ^{k}) \right \|_{2}^{2} + \frac{\alpha}{2\rho_{1}}\left \| \sum_{m=1}^{M}\varepsilon_{b_{max}}(\hat{\boldsymbol{g}}_{m}^{k}) \right \|_{2}^{2}  + (\frac{L}{2}+\beta_{1})(1+\rho _{2}) \alpha^{2} \left \| \sum_{m=1}^{M}Q_{b_{max}}(\hat{\boldsymbol{g}}_{m}^{k}) \right \|_{2}^{2} \nonumber \\
&+\sum_{d=1}^{D-1}(\beta_{d+1}-\beta_{d})\left \| \boldsymbol{\theta} ^{k+1-d}- \boldsymbol{\theta} ^{k-d} \right \|_{2}^{2}-\beta_{D}\left \| \boldsymbol{\theta} ^{k+1-D}- \boldsymbol{\theta} ^{k-D} \right \|_{2}^{2} \nonumber \\
& +  [\frac{\alpha}{2}+(\frac{L}{2}+\beta_{1})(1+\rho _{2}^{-1}) \alpha^{2}]\frac{1}{\alpha ^{2}}\sum_{d=1}^{D}\xi _{d}\left \| \boldsymbol{\theta} ^{k+1-d}-\boldsymbol{\theta} ^{k-d} \right \|_{2}^{2}  \nonumber \\ 
& + 3[\frac{\alpha}{2}+(\frac{L}{2}+\beta_{1})(1+\rho _{2}^{-1}) \alpha^{2}]M\sum_{m\in \mathbb{M}_{0}^{k}}(\left \| \varepsilon _{b_{max}}(\hat{\boldsymbol{g}}_{m}^{k-1})\right \|_{2}^{2}+\left \| \varepsilon _{b_{max}}({\boldsymbol{g}}_{m}^{k}) \right \|_{2}^{2}) \nonumber \\
& +  2[\frac{\alpha}{2}+(\frac{L}{2}+\beta_{1})(1+\rho _{2}^{-1}) \alpha^{2}]M( \sum_{b=1}^{b_{max}}\sum_{m\in \mathbb{M}_{b}^{k}} \left \| \varepsilon _{\hat{b}_{m}^{k}}(\hat{\boldsymbol{g}}_{m}^{k}) \right \|_{2}^{2} +  \sum_{b=1}^{b_{max}}\sum_{m\in \mathbb{M}_{b}^{k}} \left \| \varepsilon _{b_{max}}(\hat{\boldsymbol{g}}_{m}^{k}) \right \|_{2}^{2})  \\ 
&= (-\frac{\alpha}{2}+\frac{\alpha\rho_{1}}{2})\left \| \nabla f(\boldsymbol{\theta} ^{k}) \right \|_{2}^{2} + \frac{\alpha}{2\rho_{1}}\left \| \sum_{m=1}^{M}\varepsilon_{b_{max}}(\hat{\boldsymbol{g}}_{m}^{k}) \right \|_{2}^{2}  + (\frac{L}{2}+\beta_{1})(1+\rho _{2}) \alpha^{2} \left \| \nabla f(\boldsymbol{\theta} ^{k})-\sum_{m=1}^{M}\varepsilon_{b_{max}}(\hat{\boldsymbol{g}}_{m}^{k}) \right \|_{2}^{2} \nonumber \\
&+\sum_{d=1}^{D-1}(\beta_{d+1}-\beta_{d})\left \| \boldsymbol{\theta} ^{k+1-d}- \boldsymbol{\theta} ^{k-d} \right \|_{2}^{2}-\beta_{D}\left \| \boldsymbol{\theta} ^{k+1-D}- \boldsymbol{\theta} ^{k-D} \right \|_{2}^{2} \nonumber \\
& +  [\frac{\alpha}{2}+(\frac{L}{2}+\beta_{1})(1+\rho _{2}^{-1}) \alpha^{2}]\frac{1}{\alpha ^{2}}\sum_{d=1}^{D}\xi _{d}\left \| \boldsymbol{\theta} ^{k+1-d}-\boldsymbol{\theta} ^{k-d} \right \|_{2}^{2}  \nonumber \\ 
& + 3[\frac{\alpha}{2}+(\frac{L}{2}+\beta_{1})(1+\rho _{2}^{-1}) \alpha^{2}]M\sum_{m\in \mathbb{M}_{0}^{k}}(\left \| \varepsilon _{b_{max}}(\hat{\boldsymbol{g}}_{m}^{k-1})\right \|_{2}^{2}+\left \| \varepsilon _{b_{max}}({\boldsymbol{g}}_{m}^{k}) \right \|_{2}^{2}) \nonumber \\
& +  2[\frac{\alpha}{2}+(\frac{L}{2}+\beta_{1})(1+\rho _{2}^{-1}) \alpha^{2}]M( \sum_{b=1}^{b_{max}}\sum_{m\in \mathbb{M}_{b}^{k}} \left \| \varepsilon _{\hat{b}_{m}^{k}}(\hat{\boldsymbol{g}}_{m}^{k}) \right \|_{2}^{2} +  \sum_{b=1}^{b_{max}}\sum_{m\in \mathbb{M}_{b}^{k}} \left \| \varepsilon _{b_{max}}(\hat{\boldsymbol{g}}_{m}^{k}) \right \|_{2}^{2})  \\ 
&\leq [-\frac{\alpha}{2}+\frac{\alpha\rho_{1}}{2}+(L+2\beta_{1})(1+\rho _{2})\alpha^2]\left \| \nabla f(\boldsymbol{\theta} ^{k}) \right \|_{2}^{2} + [\frac{\alpha}{2\rho_{1}}+(L+2\beta_{1})(1+\rho _{2})\alpha^2]\left \| \sum_{m=1}^{M}\varepsilon_{b_{max}}(\hat{\boldsymbol{g}}_{m}^{k}) \right \|_{2}^{2}  \nonumber \\
& +  \left \{[\frac{\alpha}{2}+(\frac{L}{2}+\beta_{1})(1+\rho _{2}^{-1}) \alpha^{2}]\frac{1}{\alpha ^{2}}\xi _{D}-\beta{D}\right \} \left \| \boldsymbol{\theta} ^{k+1-D}- \boldsymbol{\theta} ^{k-D} \right \|_{2}^{2}\nonumber \\
&+ \sum_{d=1}^{D-1} \left \{[\frac{\alpha}{2}+(\frac{L}{2}+\beta_{1})(1+\rho _{2}^{-1}) \alpha^{2}]\frac{1}{\alpha ^{2}}\xi _{d}+\beta_{d+1}-\beta{d}\right \}\left \| \boldsymbol{\theta} ^{k+1-d}-\boldsymbol{\theta} ^{k-d} \right \|_{2}^{2}  \nonumber \\ 
& + 3[\frac{\alpha}{2}+(\frac{L}{2}+\beta_{1})(1+\rho _{2}^{-1}) \alpha^{2}]M\sum_{m\in \mathbb{M}_{0}^{k}}(\left \| \varepsilon _{b_{max}}(\hat{\boldsymbol{g}}_{m}^{k-1})\right \|_{2}^{2}+\left \| \varepsilon _{b_{max}}({\boldsymbol{g}}_{m}^{k}) \right \|_{2}^{2}) \nonumber \\
& +  2[\frac{\alpha}{2}+(\frac{L}{2}+\beta_{1})(1+\rho _{2}^{-1}) \alpha^{2}]M( \sum_{b=1}^{b_{max}}\sum_{m\in \mathbb{M}_{b}^{k}} \left \| \varepsilon _{\hat{b}_{m}^{k}}(\hat{\boldsymbol{g}}_{m}^{k}) \right \|_{2}^{2} +  \sum_{b=1}^{b_{max}}\sum_{m\in \mathbb{M}_{b}^{k}} \left \| \varepsilon _{b_{max}}(\hat{\boldsymbol{g}}_{m}^{k}) \right \|_{2}^{2}) 
\end{align}

Ignoring the quantization errors, the following three inequalities should hold simultaneously for $\forall d \in \{1,...,D\}$ in order to ensure $\mathbb{V}(\boldsymbol{\theta} ^{k+1})-\mathbb{V}(\boldsymbol{\theta} ^{k}) \leq 0$:
\begin{subequations}
	\begin{align}
	-\frac{\alpha }{2}+\frac{1}{2}\alpha \rho _{1}+(L+2\beta _{1})(1+\rho _{2})\alpha ^{2} \leq 0 \label{ZZZa}\\
	[\frac{\alpha }{2}+(\frac{L}{2}+\beta _{1})(1+{\rho _{2}}^{-1})\alpha ^{2}]\frac{\xi _{D}}{\alpha^{2}}-\beta _{D} \leq 0 \label{ZZZb} \\
	[\frac{\alpha }{2}+(\frac{L}{2}+\beta _{1})(1+{\rho _{2}}^{-1})\alpha ^{2}]\frac{\xi _{d}}{\alpha^{2}}+\beta _{d+1}-\beta _{d} \leq 0 \label{ZZZc}
	\end{align}
	\label{eq:lemma12}
\end{subequations}

\eqref{eq:lemma12} provides the choice of range in terms of stepsize $\alpha$ and weights $\left \{ \xi_{d} \right \}_{d=1}^{D}$:

\begin{subequations}
	\begin{align}
	\sum_{d=1}^{D}\xi_{d} &\leq \min \left \{  \frac{1-\rho_{1}}{4(1+\rho_{2})}, \frac{1}{2(1+\rho_{2}^{-1})}\right \} \label{ZZZZa}\\
	\alpha &\leq \min \left \{ \frac{2}{L}\left [ \frac{1-\rho_{1}}{4(1+\rho_{2})}-\sum_{d=1}^{D}\xi_{d}\right ], \frac{2}{L}\left [\frac{1}{2(1+\rho_{2}^{-1})}-\sum_{d=1}^{D}\xi_{d}\right ]\right \} \label{ZZZZb} 
	\end{align}
\end{subequations}

The above analysis indicates that there is no need to modify these two parameters involved in LAQ\cite{LAQTPAMI}.

\subsection{Proof of Lemma 2}

Under Assumption 2:
\begin{align}
&\mathbb{V}(\boldsymbol{\theta} ^{k+1})-\mathbb{V}(\boldsymbol{\theta} ^{k}) \nonumber \\
&\leq 2\mu[-\frac{\alpha}{2}+\frac{\alpha\rho_{1}}{2}+(L+2\beta_{1})(1+\rho _{2})\alpha^2]\left [ f(\boldsymbol{\theta} ^{k})-f(\boldsymbol{\theta} ^{*}) \right ]  \nonumber \\
&+ [\frac{\alpha}{2\rho_{1}}+(L+2\beta_{1})(1+\rho _{2})\alpha^2]\left \| \sum_{m=1}^{M}\varepsilon_{b_{max}}(\hat{\boldsymbol{g}}_{m}^{k}) \right \|_{2}^{2}  \nonumber \\
& +  \beta_{D}\left \{[\frac{\alpha}{2}+(\frac{L}{2}+\beta_{1})(1+\rho _{2}^{-1}) \alpha^{2}]\frac{\xi _{D}}{\alpha ^{2}\beta_{D}}-1\right \} \left \| \boldsymbol{\theta} ^{k+1-D}- \boldsymbol{\theta} ^{k-D} \right \|_{2}^{2}\nonumber \\
&+ \sum_{d=1}^{D-1} \beta_{d}\left \{[\frac{\alpha}{2}+(\frac{L}{2}+\beta_{1})(1+\rho _{2}^{-1}) \alpha^{2}]\frac{\xi _{d}}{\alpha ^{2}\beta_{d}}+\frac{\beta_{d+1}}{\beta_{d}}-1\right \}\left \| \boldsymbol{\theta} ^{k+1-d}-\boldsymbol{\theta} ^{k-d} \right \|_{2}^{2}  \nonumber \\ 
& + 3[\frac{\alpha}{2}+(\frac{L}{2}+\beta_{1})(1+\rho _{2}^{-1}) \alpha^{2}]M\sum_{m\in \mathbb{M}_{0}^{k}}(\left \| \varepsilon _{b_{max}}(\hat{\boldsymbol{g}}_{m}^{k-1})\right \|_{2}^{2}+\left \| \varepsilon _{b_{max}}({\boldsymbol{g}}_{m}^{k}) \right \|_{2}^{2}) \nonumber \\
& +  2[\frac{\alpha}{2}+(\frac{L}{2}+\beta_{1})(1+\rho _{2}^{-1}) \alpha^{2}]M( \sum_{b=1}^{b_{max}}\sum_{m\in \mathbb{M}_{b}^{k}} \left \| \varepsilon _{\hat{b}_{m}^{k}}(\hat{\boldsymbol{g}}_{m}^{k}) \right \|_{2}^{2} +  \sum_{b=1}^{b_{max}}\sum_{m\in \mathbb{M}_{b}^{k}} \left \| \varepsilon _{b_{max}}(\hat{\boldsymbol{g}}_{m}^{k}) \right \|_{2}^{2}) 
\end{align}

Let $c$ and $B$ be defined as:
\begin{subequations}
	\begin{align}
	c = \min_{d=1,..., D} \{ &2\mu[\frac{\alpha}{2}-\frac{\alpha\rho_{1}}{2}-(L+2\beta_{1})(1+\rho _{2})\alpha^2],1 - [\frac{\alpha}{2}+(\frac{L}{2}+\beta_{1})(1+\rho _{2}^{-1}) \alpha^{2}]\frac{\xi _{D}}{\alpha ^{2}\beta_{D}},\nonumber \\ 
	&1-[\frac{\alpha}{2}+(\frac{L}{2}+\beta_{1})(1+\rho _{2}^{-1}) \alpha^{2}]\frac{\xi _{d}}{\alpha ^{2}\beta_{d}}+\frac{\beta_{d+1}}{\beta_{d}}  \} \label{ZZZZa}\\
	B=\max&\left \{ \frac{\alpha}{2\rho_{1}}+(L+2\beta_{1})(1+\rho _{2})\alpha^2, 3M[\frac{\alpha}{2}+(\frac{L}{2}+\beta_{1})(1+\rho _{2}^{-1}) \alpha^{2}] \right \} \label{ZZZZb} 
	\end{align}
\end{subequations}

Then:
\begin{align}
\mathbb{V}(\boldsymbol{\theta} ^{k+1})-\mathbb{V}(\boldsymbol{\theta} ^{k}) &\leq -c\left [ f(\boldsymbol{\theta} ^{k})-f(\boldsymbol{\theta} ^{*})+ \sum_{d=1}^{D} \beta_{d}\left \| \boldsymbol{\theta} ^{k+1-d}-\boldsymbol{\theta} ^{k-d} \right \|_{2}^{2} \right ]  \nonumber \\
&+ B \left \| \sum_{m=1}^{M}\varepsilon_{b_{max}}(\hat{\boldsymbol{g}}_{m}^{k}) \right \|_{2}^{2} +B\sum_{m\in \mathbb{M}_{0}^{k}}(\left \| \varepsilon _{b_{max}}(\hat{\boldsymbol{g}}_{m}^{k-1})\right \|_{2}^{2}+\left \| \varepsilon _{b_{max}}({\boldsymbol{g}}_{m}^{k}) \right \|_{2}^{2}) \nonumber \\
&+B( \sum_{b=1}^{b_{max}}\sum_{m\in \mathbb{M}_{b}^{k}} \left \| \varepsilon _{\hat{b}_{m}^{k}}(\hat{\boldsymbol{g}}_{m}^{k}) \right \|_{2}^{2} +  \sum_{b=1}^{b_{max}}\sum_{m\in \mathbb{M}_{b}^{k}} \left \| \varepsilon _{b_{max}}(\hat{\boldsymbol{g}}_{m}^{k}) \right \|_{2}^{2}) \\
&= -c\; \mathbb{V}(\boldsymbol{\theta} ^{k})  \nonumber \\
&+ B \left \| \sum_{m=1}^{M}\varepsilon_{b_{max}}(\hat{\boldsymbol{g}}_{m}^{k}) \right \|_{2}^{2} +B\sum_{m\in \mathbb{M}_{0}^{k}}(\left \| \varepsilon _{b_{max}}(\hat{\boldsymbol{g}}_{m}^{k-1})\right \|_{2}^{2}+\left \| \varepsilon _{b_{max}}({\boldsymbol{g}}_{m}^{k}) \right \|_{2}^{2}) \nonumber \\
&+B( \sum_{b=1}^{b_{max}}\sum_{m\in \mathbb{M}_{b}^{k}} \left \| \varepsilon _{\hat{b}_{m}^{k}}(\hat{\boldsymbol{g}}_{m}^{k}) \right \|_{2}^{2} +  \sum_{b=1}^{b_{max}}\sum_{m\in \mathbb{M}_{b}^{k}} \left \| \varepsilon _{b_{max}}(\hat{\boldsymbol{g}}_{m}^{k}) \right \|_{2}^{2})  
\end{align}

Thus,
\begin{align}
&\mathbb{V}(\boldsymbol{\theta} ^{k+1}) \leq (1-c)\mathbb{V}(\boldsymbol{\theta} ^{k})  \nonumber \\
&+ B \left \| \sum_{m=1}^{M}\varepsilon_{b_{max}}(\hat{\boldsymbol{g}}_{m}^{k}) \right \|_{2}^{2} +B\sum_{m\in \mathbb{M}_{0}^{k}}(\left \| \varepsilon _{b_{max}}(\hat{\boldsymbol{g}}_{m}^{k-1})\right \|_{2}^{2}+\left \| \varepsilon _{b_{max}}({\boldsymbol{g}}_{m}^{k}) \right \|_{2}^{2}) \nonumber \\
&+B( \sum_{b=1}^{b_{max}}\sum_{m\in \mathbb{M}_{b}^{k}} \left \| \varepsilon _{\hat{b}_{m}^{k}}(\hat{\boldsymbol{g}}_{m}^{k}) \right \|_{2}^{2} +  \sum_{b=1}^{b_{max}}\sum_{m\in \mathbb{M}_{b}^{k}} \left \| \varepsilon _{b_{max}}(\hat{\boldsymbol{g}}_{m}^{k}) \right \|_{2}^{2}) 
\end{align}

\subsection{Proof of Theorem 1}
This part proves that \eqref{L} holds for any $k\geq0$ if the following inequalities are satisfied:
\begin{subequations}
	\begin{align}
	& 4BMP\tau _{b_{max}}^{2} + BMP\sum_{b=1}^{b_{max}}\tau _{b_{m}^{k}}^{2} \leq \sigma_{2}-\sigma_{1} \label{ZZZZZa}\\
	&\frac{24L^{2}}{\mu} + 18\tau _{b_{max}-b_{m}^{k}}^{2} + 3\tau _{b_{max}}^{2} \leq \sigma_{2}
	\label{ZZZZZb} \\
	&\alpha \geq \frac{\mu}{4L^{2}M^{2}} \label{ZZZZZc} 
	\end{align}
\end{subequations}

It is assumed that for any
$k \geq 1$, \eqref{L} holds for $k-1$. Let $\sigma_{1}=1-c$, there is:
\begin{align}
&\mathbb{V}(\boldsymbol{\theta} ^{k+1}) \leq \sigma_{1}\mathbb{V}(\boldsymbol{\theta} ^{k})  \nonumber \\
&+ B \left \| \sum_{m=1}^{M}\varepsilon_{b_{max}}(\hat{\boldsymbol{g}}_{m}^{k}) \right \|_{2}^{2} +B\sum_{m\in \mathbb{M}_{0}^{k}}(\left \| \varepsilon _{b_{max}}(\hat{\boldsymbol{g}}_{m}^{k-1})\right \|_{2}^{2}+\left \| \varepsilon _{b_{max}}({\boldsymbol{g}}_{m}^{k}) \right \|_{2}^{2}) \nonumber \\
&+B( \sum_{b=1}^{b_{max}}\sum_{m\in \mathbb{M}_{b}^{k}} \left \| \varepsilon _{\hat{b}_{m}^{k}}(\hat{\boldsymbol{g}}_{m}^{k}) \right \|_{2}^{2} +  \sum_{b=1}^{b_{max}}\sum_{m\in \mathbb{M}_{b}^{k}} \left \| \varepsilon _{b_{max}}(\hat{\boldsymbol{g}}_{m}^{k}) \right \|_{2}^{2})  \\
&\leq \sigma_{1}\sigma _{2}^{k-1}\mathbb{V}(\boldsymbol{\theta} ^{1}) + 4BMP\tau _{b_{max}}^{2}\sigma _{2}^{k-1}\mathbb{V}(\boldsymbol{\theta} ^{1}) + BMP\sum_{b=1}^{b_{max}}\tau _{b_{m}^{k}}^{2}\sigma _{2}^{k-1}\mathbb{V}(\boldsymbol{\theta} ^{1}) \nonumber \\
&= (\sigma_{1} + 4BMP\tau _{b_{max}}^{2} + BMP\sum_{b=1}^{b_{max}}\tau _{b_{m}^{k}}^{2})\sigma _{2}^{k-1}\mathbb{V}(\boldsymbol{\theta} ^{1}) \leq \sigma _{2}^{k}\mathbb{V}(\boldsymbol{\theta} ^{1})
\end{align}
where $\sigma _{2} \geq \sigma_{1} + 4BMP\tau _{b_{max}}^{2} + BMP\sum_{b=1}^{b_{max}}\tau _{b_{m}^{k}}^{2}$.

Under Assumption 1 and Assumption 2, the following inequality holds for any $\boldsymbol{\theta} _{1}$ and $\boldsymbol{\theta} _{2}$ because of convexity:

\begin{align}
\left \| \nabla f_{m}(\boldsymbol{\theta} _{1}) - \nabla f_{m}(\boldsymbol{\theta} _{2})\right \|_{\infty}
&\leq \left \| \sum_{m=1}^{M}(\nabla f_{m}(\boldsymbol{\theta} _{1})-\nabla f_{m}(\boldsymbol{\theta} _{2})) \right \|_{\infty} \nonumber \\
&=\left \| \nabla f(\boldsymbol{\theta} _{1}) - \nabla f(\boldsymbol{\theta} _{2})\right \|_{\infty} \nonumber \\
&\leq L\left \|  \boldsymbol{\theta} _{1} -  \boldsymbol{\theta} _{2}\right \|_{\infty}, \; \forall m \in \{1,...,M\}
\label{eq:42}
\end{align}

With \eqref{eq:42} and the proposed precision selection criterion \eqref{eq6}, there is:
\begin{align}
&\left \| \nabla f_{m}(\boldsymbol{\theta} ^{k+1}) - Q_{\hat{b}_{m}^{k-1}}(\hat{\boldsymbol{g}}_{m}^{k-1})\right \|_{\infty} \nonumber \\
&=\left \| \nabla f_{m}(\boldsymbol{\theta} ^{k+1}) - f_{m}(\boldsymbol{\theta} ^{k}) + f_{m}(\boldsymbol{\theta} ^{k}) - Q_{b_{max}}(\boldsymbol{\boldsymbol{g}}_{m}^{k}) + Q_{b_{max}}(\boldsymbol{\boldsymbol{g}}_{m}^{k}) - Q_{\hat{b}_{m}^{k-1}}(\hat{\boldsymbol{g}}_{m}^{k-1})\right \|_{\infty} \nonumber \\
&\leq \left \| \nabla f_{m}(\boldsymbol{\theta} ^{k+1}) - f_{m}(\boldsymbol{\theta} ^{k})\right \|_{\infty} + \left \|  f_{m}(\boldsymbol{\theta} ^{k}) - Q_{b_{max}}(\boldsymbol{g}_{m}^{k})\right \|_{\infty} + \left \|  Q_{b_{max}}(\boldsymbol{g}_{m}^{k}) - Q_{\hat{b}_{m}^{k-1}}(\hat{\boldsymbol{g}}_{m}^{k-1})\right \|_{\infty} \nonumber \\
&\leq L\left \| \boldsymbol{\theta} ^{k+1}- \boldsymbol{\theta} ^{k} \right \|_{\infty} + \left \|\varepsilon _{b_{max}}(\boldsymbol{g}_{m}^{k})\right \|_{\infty} \nonumber \\
&+\sqrt{\frac{1}{\alpha ^{2}M^{2}}\sum_{d=1}^{D}\xi _{d}\left \| \boldsymbol{\theta} ^{k+1-d}-\boldsymbol{\theta} ^{k-d} \right \|_{2}^{2} +3(\left \| \varepsilon _{b_{max}-b_{m}^{k}}(\hat{\boldsymbol{g}}_{m}^{k-1})\right \|_{2}^{2}+\left \| \varepsilon _{b_{max}-b_{m}^{k}}({\boldsymbol{g}}_{m}^{k}) \right \|_{2}^{2})}  \\
&\leq L\sqrt{\left \| \boldsymbol{\theta} ^{k+1}-\boldsymbol{\theta} ^{*}+\boldsymbol{\theta} ^{*}- \boldsymbol{\theta} ^{k} \right \|_{2}^{2}} + \left \|\varepsilon _{b_{max}}(\boldsymbol{g}_{m}^{k})\right \|_{\infty} \nonumber \\
&+\sqrt{\frac{1}{\alpha ^{2}M^{2}}\sum_{d=1}^{D}\xi _{d}\left \| \boldsymbol{\theta} ^{k+1-d}-\boldsymbol{\theta} ^{k-d} \right \|_{2}^{2} +3(\left \| \varepsilon _{b_{max}-b_{m}^{k}}(\hat{\boldsymbol{g}}_{m}^{k-1})\right \|_{2}^{2}+\left \| \varepsilon _{b_{max}-b_{m}^{k}}({\boldsymbol{g}}_{m}^{k}) \right \|_{2}^{2})}  \\
&\leq L\sqrt{2\left \| \boldsymbol{\theta} ^{k+1}-\boldsymbol{\theta} ^{*} \right \|_{2}^{2}+2\left \| \boldsymbol{\theta} ^{*}- \boldsymbol{\theta} ^{k} \right \|_{2}^{2}} + \left \|\varepsilon _{b_{max}}(\boldsymbol{g}_{m}^{k})\right \|_{\infty} \nonumber \\
&+\sqrt{\frac{1}{\alpha ^{2}M^{2}}\sum_{d=1}^{D}\xi _{d}\left \| \boldsymbol{\theta} ^{k+1-d}-\boldsymbol{\theta} ^{k-d} \right \|_{2}^{2} +3(\left \| \varepsilon _{b_{max}-b_{m}^{k}}(\hat{\boldsymbol{g}}_{m}^{k-1})\right \|_{\infty}^{2}+\left \| \varepsilon _{b_{max}-b_{m}^{k}}({\boldsymbol{g}}_{m}^{k}) \right \|_{\infty}^{2})}
\end{align}

Under Assumption 2 with \eqref{eq11},

\begin{align}
&\left \| \nabla f_{m}(\boldsymbol{\theta} ^{k+1}) - Q_{\hat{b}_{m}^{k-1}}(\hat{\boldsymbol{g}}_{m}^{k-1})\right \|_{\infty}^{2} \nonumber \\
&\leq \frac{12L^{2}}{\mu}\left [ f(\boldsymbol{\theta} ^{k+1})-f(\boldsymbol{\theta} ^{*})+f(\boldsymbol{\theta} ^{k})-f(\boldsymbol{\theta} ^{*}) \right ]+ 3\left \|\varepsilon _{b_{max}}(\boldsymbol{g}_{m}^{k})\right \|_{\infty}^{2} \nonumber \\
& + \frac{3}{\alpha ^{2}M^{2}}\sum_{d=1}^{D}\xi _{d}\left \| \boldsymbol{\theta} ^{k+1-d}-\boldsymbol{\theta} ^{k-d} \right \|_{2}^{2} +9(\left \| \varepsilon _{b_{max}-b_{m}^{k}}(\hat{\boldsymbol{g}}_{m}^{k-1})\right \|_{\infty}^{2}+\left \| \varepsilon _{b_{max}-b_{m}^{k}}({\boldsymbol{g}}_{m}^{k}) \right \|_{\infty}^{2})  \\
& \leq \frac{12L^{2}}{\mu}\left [ f(\boldsymbol{\theta} ^{k+1})-f(\boldsymbol{\theta} ^{*})+f(\boldsymbol{\theta} ^{k})-f(\boldsymbol{\theta} ^{*}) + \frac{\mu}{4L^{2}\alpha ^{2}M^{2}}\sum_{d=1}^{D}\xi _{d}\left \| \boldsymbol{\theta} ^{k+1-d}-{\boldsymbol{\theta}} ^{k-d} \right \|_{2}^{2} \right ] \nonumber \\
& + 18P\tau _{b_{max}-b_{m}^{k}}^{2}\sigma _{2}^{k-1}\mathbb{V}({\boldsymbol{\theta}} ^{1}) + 3P\tau _{b_{max}}^{2}\sigma _{2}^{k-1}\mathbb{V}({\boldsymbol{\theta}} ^{1})
\end{align}

With $\alpha \geq \frac{\mu}{4L^{2}M^{2}}$, $ \frac{\mu\xi_{d}}{4L^{2}\alpha ^{2}M^{2}} \leq \frac{\xi_{d}}{\alpha} \leq \sum_{j=d}^{D}\frac{\xi_{j}}{\alpha}$:

\begin{align}
&\left \| \nabla f_{m}({\boldsymbol{\theta}} ^{k+1}) - Q_{\hat{b}_{m}^{k-1}}(\hat{\boldsymbol{g}}_{m}^{k-1})\right \|_{\infty}^{2} \nonumber \\
& \leq \frac{12L^{2}}{\mu}\left [ f({\boldsymbol{\theta}} ^{k+1})-f({\boldsymbol{\theta}} ^{*})+f({\boldsymbol{\theta}} ^{k})-f({\boldsymbol{\theta}} ^{*}) + \sum_{d=1}^{D}\sum_{j=d}^{D}\frac{\xi_{j}}{\alpha}\left \| {\boldsymbol{\theta}} ^{k+1-d}-{\boldsymbol{\theta}} ^{k-d} \right \|_{2}^{2} \right ] \nonumber \\
& + 18P\tau _{b_{max}-b_{m}^{k}}^{2}\sigma _{2}^{k-1}\mathbb{V}({\boldsymbol{\theta}} ^{1}) + 3P\tau _{b_{max}}^{2}\sigma _{2}^{k-1}\mathbb{V}({\boldsymbol{\theta}} ^{1}) \nonumber \\
& \leq \frac{12L^{2}}{\mu}\left [ \mathbb{V}({\boldsymbol{\theta}} ^{k+1}) + \mathbb{V}({\boldsymbol{\theta}} ^{k}) \right ] + 18P\tau _{b_{max}-b_{m}^{k}}^{2}\sigma _{2}^{k-1}\mathbb{V}({\boldsymbol{\theta}} ^{1}) + 3P\tau _{b_{max}}^{2}\sigma _{2}^{k-1}\mathbb{V}({\boldsymbol{\theta}} ^{1}) \nonumber \\
& \leq \frac{24L^{2}}{\mu}\sigma _{2}^{k-1}\mathbb{V}({\boldsymbol{\theta}} ^{1}) + 18P\tau _{b_{max}-b_{m}^{k}}^{2}\sigma _{2}^{k-1}\mathbb{V}({\boldsymbol{\theta}} ^{1}) + 3P\tau _{b_{max}}^{2}\sigma _{2}^{k-1}\mathbb{V}({\boldsymbol{\theta}} ^{1}) \nonumber \\
&= (\frac{24L^{2}}{\mu P} + 18\tau _{b_{max}-b_{m}^{k}}^{2} + 3\tau _{b_{max}}^{2})P\sigma _{2}^{k-1}\mathbb{V}({\boldsymbol{\theta}} ^{1}) \leq P\sigma _{2}^{k}\mathbb{V}({\boldsymbol{\theta}} ^{1})
\end{align}

Thus,
\begin{align}
\left \| \varepsilon _{b}(\boldsymbol{g}_{m} ^{k}) \right \|_{\infty}^{2}\leq {\tau_{b}}^{2} \left \| \nabla f_{m}({\boldsymbol{\theta}} ^{k+1}) - Q_{\hat{b}_{m}^{k-1}}(\hat{\boldsymbol{g}}_{m}^{k-1})\right \|_{\infty}^{2} \leq P{\tau_{b}}^{2} \sigma _{2}^{k}\mathbb{V}({\boldsymbol{\theta}} ^{1})
\end{align}

\section{Simulation datasets}

\begin{table}[htbp]
	\caption{The heterogeneous simulation datasets used for logistic regression.}
	\begin{tabular}{c|c|c|c}
		\hline
		Dataset    & \# features & \# samples & client index      \\ \hline
		Adult fat\cite{1997Scaling}  & 113        & 1605      & 1,2,3,4,5,6       \\
		Ionosphere\cite{1989Classification} & 34         & 351       & 7,8,9,10,11,12    \\
		Derm\cite{HA1998Learning}       & 34         & 358       & 13,14,15,16,17,18 \\ \hline
	\end{tabular}
    \label{tb:dataset}
\end{table}

Three binary classification datasets listed in Table.~\ref{tb:dataset} are used together in order to simulate non-IID data distribution as Chen \textit{et al.} do in the evaluation of LAQ \cite{chen2018lag}. Specifically, 
The number of features is preprocessed to be equal to the minimal number of features among the total three datasets, and each dataset is uniformly distributed across six clients.

\end{document}